\documentclass[a4paper,twocolumn,10pt,accepted=2025-08-06]{quantumarticle}
\pdfoutput=1
\usepackage[numbers,sort&compress]{natbib}
\usepackage[utf8]{inputenc}
\usepackage[english]{babel}
\usepackage[T1]{fontenc}
\usepackage{soul}

\usepackage{hyperref}
\hypersetup{
	colorlinks,
	linkcolor={blue},
	citecolor={blue},
	urlcolor={blue}
}
\usepackage{bm}
\usepackage{epsfig}
\usepackage{amssymb}
\usepackage{graphicx}
\usepackage{mathrsfs}
\usepackage{amsmath}
\usepackage{marvosym}
\usepackage{bbm}
\usepackage{bbold}
\usepackage[mathscr]{eucal}
\usepackage{color}
\usepackage{multirow}
\usepackage{adjustbox}
\usepackage{tikz}
\usetikzlibrary{arrows,%
                shapes,positioning}

\usepackage{lipsum}
\usepackage[scr=rsfs]{mathalfa}
\newcommand{\centered}[1]{\begin{tabular}{l} #1 \end{tabular}}
\newcommand{\LL}{\ell}

\usepackage{array, makecell}
\usepackage{placeins}

\begin{document}

\title{Minimal criteria for continuous-variable genuine multipartite entanglement}


\author{Olga Leskovjanov{\'a}}
\email{ola.leskovjanova@gmail.com}
\orcid{0000-0002-3638-3627}
\affiliation{Department of Optics, Palack\'y University, 17. listopadu 12, 771~46 Olomouc, Czech Republic}

\author{Ladislav Mi\v{s}ta, Jr.}
\email{mista@optics.upol.cz}
\orcid{0000-0002-1979-7617}
\affiliation{Department of Optics, Palack\'y University, 17. listopadu 12, 771~46 Olomouc, Czech Republic}

\maketitle


\begin{abstract}
We derive a set of genuine multi-mode entanglement criteria for second moments of the quadrature operators. The criteria have a common
form of the uncertainty relation between sums of variances of position and momentum quadrature combinations. A unique
feature of the criteria is that the sums contain the least possible number of variances of at most two-mode combinations.
The number of second moments we need to know to apply the criteria thus scales only linearly with the number of modes,
as opposed to the quadratic scaling of the already existing criteria. Each criterion is associated with a tree graph, which allowed
us to develop a direct method of construction of the criteria based solely on the structure of the underlying tree.
The practicality of the proposed criteria is demonstrated by finding a number of examples of Gaussian states of up to six modes,
whose genuine multi-mode entanglement is detected by them. The designed criteria are particularly suitable for verification of
genuine multipartite entanglement in large multi-mode states or when only a set of two-mode nearest-neighbour marginal
covariance matrices of the investigated state is available.
\end{abstract}


\section{Introduction}\label{sec_Introduction}

At the very foundations of the quantum information theory lies the effort to explain possible relationships
between a pair of quantum systems. These relationships manifest itself through various forms of
correlations which we try to capture by intuitive models complying better with our idea of how the nature ought to work. Then we can classify
states of a composite system into a nested family of sets according to whether the state admits a separable model \cite{Werner_89,Horodecki_09},
local hidden state model \cite{Wiseman_07,Uola_20} or local hidden variable model \cite{Bell_64,Brunner_14}. The absence of the model then
points at a closer relationship between the subsystems, which is referred to as entanglement, steering and non-locality,
and which can be used as a resource for quantum communication and computing \cite{Nielsen_00}.

The complexity of the relationships is significantly increased in systems consisting of three or more subsystems,
called commonly as the multipartite systems. From the point of view of quantum entanglement, states of multipartite systems can be classified into two disjoint sets. One is the set of so-called biseparable states, which can be prepared by statistical mixing of states that are separable with respect to (generally different) partitions of all subsystems into two groups. The second is then the set of states that cannot be prepared in this way, which we refer to as genuine multipartite entangled (GME) \cite{Bancal_11}.

Genuine multiparticle entanglement is of great importance both in terms of fundamental research and applications. The GME carried by the Greenberger-Horne-Zeilinger (GHZ) state exhibits a strong contradiction with locally realistic theories, which can be captured without the use of the Bell-type statistical inequalities \cite{Greenberger_89,Pan_00}. The same state also serves as an ideal testbed for the analysis of the scaling law of the coherence decay of a multiparticle quantum register \cite{Monz_11}. Early applications of GME included quantum secret sharing protocol \cite{Hillery_99} or assisted quantum teleportation schemes \cite{Karlsson_98,Yonezawa_04}. In newer applications, GME is used to increase the precision of phase estimation in interferometry \cite{Giovannetti_04,Toth_12} and it also plays a key role in the measurement-based model of quantum computing \cite{Raussendorf_01,Walther_05}, in which the computation is performed using a sequence of local single-site measurements on a large GME state. Last but not least, GME is also a resource for multiparty quantum key distribution \cite{Epping_17,Ribeiro_18}. 


Motivated by both the fundamental questions of many-body physics and the desire to demonstrate the advantage of quantum
computers compared to classic ones, experimenters have been trying to create ever-larger GME states.
An important part of characterizing these states is obviously the ability to detect the GME. There are at least three reasons for this.
First, certification of GME confirms the presence of a key resource for measurement-based model of quantum computing \cite{Raussendorf_01}.
Second, detection of the GME is crucial for understanding of the entanglement dynamics in a simulated system \cite{Friis_18}.
Finally, on a general level successful generation of GME evidences a high degree of control which was gained over the used experimental
platform \cite{Gong_19}.

Attempts to prepare a large GME state are proceeding in two directions. One uses systems with a two-dimensional Hilbert state space
(qubits), while the other relies on systems with an infinite-dimensional Hilbert state space, the so-called continuous-variable (CV) systems.
Nowadays, preparation of the qubit GME states is limited to the medium-scale systems. This includes preparation of the 14 ion qubits
in the GHZ state \cite{Monz_11}, 18-qubit photonic GHZ state \cite{Wang_18} or 51-qubit cluster state
on a superconducting quantum processor \cite{Cao_23} to name at least few of them.

A more auspicious framework for preparation of a large GME state is provided by Gaussian states of CV systems \cite{Weedbrook_12}. These systems
are realized typically by optical modes and they are characterized by canonically conjugate position $x$ and momentum $p$ quadrature
operators. Gaussian states possess a Gaussian-shaped Wigner function and they are easily prepared, manipulated and measured in the laboratory.
Additionally, entanglement properties of an $N$-mode Gaussian state are completely characterized by $N(2N+1)$ independent elements of its
covariance matrix and thus the number of parameters determining the state grows quadratically with the number of modes.
Although so far only three-mode \cite{Shalm_13,Armstrong_15} and four-mode \cite{Gerke_15,Shchukin_15} Gaussian GME states have been demonstrated
in the laboratory, medium-scale to large-scale Gaussian states possessing a weaker form of global entanglement, known as the full inseparability,
have been already observed experimentally. Specifically, a fully inseparable 10-mode \cite{Gerke_15} and 60-mode \cite{Chen_14} state has
been prepared using optical frequency comb, and 10 000-mode \cite{Yokoyama_13} or even one-million-mode \cite{Yoshikawa_16} fully inseparable state
has been generated by means of the time-domain multiplexing.

There are two basic ways in which the CV GME can be detected. The first one utilizes the GME witness in the space of covariance matrices
\cite{Hyllus_06}. It requires knowledge of the entire covariance matrix, or its entire position and momentum part, and so far it
has been used exclusively for verification of the GME in the theoretical states \cite{Paternostro_07,Jolin_23}. The second approach is based
on simpler CV GME criteria \cite{Shalm_13,Teh_14,Shchukin_15,Toscano_15}, which are more transparent and more economical in terms of the number of
needed measurements. All the criteria are built on the van Loock-Furusawa criterion of full inseparability \cite{vanLoock_03}, which generalizes Duan's {\it et.\! al.}
criterion of two-mode entanglement \cite{Duan_00} and its bipartite generalization \cite{Giovannetti_03} to the multi-mode case. The considered criteria are usually designed to detect in a simple way some paradigmatic
GME Gaussian states and therefore they are of the form of a single inequality for the second moments of the quadrature operators. As an example can
serve us the sum version of the criterion \cite{Shalm_13}, according to which a three-mode state is GME if the inequality
\begin{align}\label{Shalm}
&\langle\left[\Delta\left(x_{1}-x_{2}\right)\right]^{2}\rangle+\langle\left[\Delta\left(x_{2}-x_{3}\right)\right]^2\rangle\nonumber\\
&+\langle\left[\Delta\left(p_{1}+p_{2}+p_{3}\right)\right]^2\rangle\geq2
\end{align}
is violated. The criterion exhibits a common feature of the overwhelming majority of the practically used GME criteria. Namely,
some (more often all \cite{Teh_14,Armstrong_15}) occurring variances contain a global combination of all position or momentum quadratures.
For the three-mode criterion (\ref{Shalm}) the combination in question is the total momentum quadrature $p_{1}+p_{2}+p_{3}$,
whereas more general $N$-mode criteria typically contain the weighted combination $\sum_{i=1}^{N}g_{i}p_{i}$.
Consequently, utilization of such the criterion requires knowledge of all independent elements of the momentum part of the covariance matrix,
the number of which scales quadratically with $N$. Guided by the effort to detect GME of the largest possible state, we can ask
ourselves whether this quadratic growth can be slowed down, e.g., just to the linear growth. Finding a positive answer
to this question would undoubtedly be an important result because then the GME certification of some CV states would require
a significantly smaller number of measurements. However, such a result would be rather counterintuitive because the
respective criterion then could not contain any global combination mentioned above, which might seem to be indispensable
for the detection of the global property of the GME.

Interestingly, this intuition is false. Namely, there are two similar criteria of GME,
which nevertheless contain variances of combinations of only pairs of position and momentum quadratures.
More precisely, one criterion is $N$-mode and it is given by \cite[Eq.~(25)]{Shchukin_15},
\begin{equation}\label{Shchukin}
\sum_{1\leq i<j\leq N}\langle\left(x_{i}+x_{j}\right)^{2}+\left(p_{i}-p_{j}\right)^2\rangle\geq f(N),
\end{equation}
where $f(N)$ is a function of $N$, whereas the other is four-mode and employs variances of the same
combinations (up to the sign) of quadrature operators \cite[Eqs.~(44) and (47)]{Toscano_15}. Importantly,
both the criteria contain combinations for {\it all} possible pairs of the modes. Thus as before, the number of the elements of the
covariance matrix one needs for application of both the latter criteria grows quadratically with the number of modes $N$.

Continuing the same line of thoughts, one can then ask whether the latter criteria are already simplest criteria capable
of detecting the GME or their structure can be simplified even further. It is obvious that one cannot simplify the form of the
combinations, because they are just two-mode. But since the criteria use the entire covariance matrix of the investigated state,
what could be reduced in principle is the number of appearing combinations. Recent results on witnessing qubit GME from separable
nearest-neighbour marginals \cite{Paraschiv_17,Micuda_19} and their extension to the CV systems \cite{Nordgren_22} indicate that this simplification
should be indeed possible.

In this paper we derive such the simplified CV GME criteria for $N$ modes. Similar to the known criteria
also the proposed criteria have the standard form akin to the product and sum uncertainty relation for elements of
the covariance matrix and like the criteria \cite[Eq.~(25)]{Shchukin_15} and \cite[Eqs.~(44) and (47)]{Toscano_15}
they include at most two-mode quadrature combinations. However, unlike any of the existing CV GME criteria they contain the minimum
number of the combinations. The number is minimal because it is equal to the number of elements in the minimal set
of two-mode marginal covariance matrices \cite{Paraschiv_17,Nordgren_22}, which is needed for detection of GME.
Since the number of independent covariance matrix elements contained in the minimal set is equal to $7N-4$ \cite{Nordgren_22} and thus
grows linearly with $N$, the number of elements of the covariance matrix one needs to know for application of the proposed criteria
also scales only linearly with $N$. As every minimal set can be represented by a tree graph,
each of the presented criterion corresponds to a tree. This allowed us to develop a direct method of
construction of the criteria, which is based solely on the structure of the underlying tree. We also
demonstrate practicality of our criteria by finding analytical as well as numerical examples of
Gaussian states whose GME is detected by the criteria for almost all configurations of up to six modes.
Interestingly, our criterion detects the GME of a three-mode CV GHZ-like state from its two two-mode reduced density matrices. This is in sharp contrast to the qubit GHZ state for which detection of the GME from two-qubit marginals is impossible \cite{Gittsovich_10} and illustrates another difference between discrete and continuous quantum variables.

The proposed criteria are most economic as far as the number of required measurements is concerned and thus they can serve for verification of GME in large multi-mode states. They can be used to detect CV GME even in cases where only the elements of two-mode marginal covariance matrices belonging to some minimal set are available. From a different point of view, the derived criteria can be interpreted in the context of the so-called quantum marginal problem \cite{Higuchi_03,Bravyi_03}. In its broader sense, this problem addresses the question of determining the presence of some global property based on the knowledge of only a set of reduced density matrices \cite{Wurflinger_12,Chen_14}. The task of finding the conditions under which the given reductions are compatible with the property under consideration is then known as the quantum marginal problem. From this perspective, each of the presented criteria can be viewed as a necessary condition for the corresponding minimal set of two-mode covariance matrices having a common global state to be compatible with a biseparable state.

The paper is structured as follows. In Sec.~\ref{sec_GME} we give a brief introduction into the concept of GME, whereas Sec.~\ref{sec_CV_entanglement} is dedicated to the entanglement
in CV systems. Section~\ref{sec_minimal_criteria} contains derivation of the minimal criteria for CV GME and the Gaussian GME states which are detected by the criteria are constructed in
Sec.~\ref{sec_applications}. Conclusions are in Sec.~\ref{sec_conclusions}.

\section{Genuine multipartite entanglement}\label{sec_GME}

We deal with $N$ quantum systems $1,2,\ldots,N$, which we collect into the
set $\mathcal{M}=\{1,2,\ldots,N\}$. Consider further the split of the set $\mathcal{M}$ into two disjoint
nonempty subsets $I_{k}=\{i_{1},i_{2},\ldots,i_{l}\}$, $i_{n}\in\mathcal{M}$, $n=1,2,\ldots,l$, $0<l<N$, and
$J_{k}=\mathcal{M}\setminus I_{k}$. Here, the index $k=1,2,\ldots,K_{N}$ labels all different
inequivalent splits of the set $\mathcal{M}$ and $K_{N}=2^{N-1}-1$ is the number of the splits. The splits shall be henceforth called as bipartite splits and denoted as
$I_{k}|J_{k}$. We then say that a state $\rho$ of the considered system is separable with respect to the
split $I_{k}|J_{k}$ if it can be expressed as \cite{Werner_89,Horodecki_09}
\begin{equation}\label{eq:rhosep}
\rho=\rho_{I_{k}|J_{k}}^{\rm sep}=\sum_{i}p_{i}\rho_{I_{k}}^{(i)}\otimes\rho_{J_{k}}^{(i)},
\end{equation}
where $\rho_{I_{k}}^{(i)}$ and $\rho_{J_{k}}^{(i)}$ are local states of subsystems belonging to the set $I_{k}$ and $J_{k}$, respectively,
and $p_{i}$ are probabilities. The states which cannot be expressed in the form (\ref{eq:rhosep}) are then called as
entangled with respect to the split and this sort of entanglement is referred to as the bipartite entanglement.

States of systems made up of only two subsystems ($N=2$) are either entangled or separable, but in systems consisting of
more subsystems ($N>2$) different types of entanglement exist \cite{Dur_99,Giedke_01}. Various forms  of the
multipartite entanglement can be illustrated on the system of just three subsystems $1,2$ and $3$. In this case, there are three bipartite splits
$1|23, 2|31$ and $3|12$, as well as a tripartite split $1|2|3$, and one can classify all states into five
separability classes \cite{Dur_99,Giedke_01} depending on whether they are separable with respect to tripartite split, or only with respect to none, one, two, or three bipartite splits. Clearly, from the point of view of this classification the strongest form
of multipartite entanglement is carried by the so-called fully inseparable states, which are
entangled across all three bipartite splits.

A specific feature of the most important fully inseparable states, such as the GHZ state, is that global quantum operation on all three subsystems is needed to prepare them. However, this is not a common property of all fully inseparable states. Namely, there is a class of fully inseparable states that can be prepared more easily by randomly mixing states that do not require such a global operation for their preparation \cite{Piani_07}. To show this, consider a state $|\chi\rangle_{ijk}\equiv|\phi_{+}\rangle_{ij}|0\rangle_{k}$ of three qubits $i, j, k$,  where $|\phi_{+}\rangle_{ij}=(|00\rangle_{ij}+|11\rangle_{ij})/\sqrt{2}$ is the Bell state, whose preparation obviously does not need a global three-qubit quantum operation. Let us now imagine that we randomly prepare with the same probability $1/3$ one of the states  $|\chi\rangle_{123}, |\chi\rangle_{132}$ and $|\chi\rangle_{231}$. Interestingly, the result is also a fully inseparable state \cite{Guhne_09}. To distinguish states that can be prepared by convex mixing of the product of states with respect to different bipartite splits, called biseparable states \cite{Bancal_11}, from states that cannot be prepared in this way, the concept of GME has been introduced as follows.
Generalizing the notion of a biseparable state to $N$-partite system,
\begin{equation}\label{eq:rhobisep}
\rho^{\rm bisep}=\sum_{k=1}^{K_{N}}\lambda_{k}\rho_{I_{k}|J_{k}}^{\rm sep},
\end{equation}
where $\lambda_{j}$ are probabilities, we say that a state $\rho$ is GME,
if it cannot be expressed in the form (\ref{eq:rhobisep}).

Genuine multipartite entanglement finds applications in many branches of quantum information science.
For this reason, it is desirable to develop efficient GME criteria. Because the set of
biseparable states is convex, we can use the machinery of entanglement witnesses \cite{Horodecki_96,Terhal_00}
for this purpose. In the next section we give a brief introduction into the GME of states
of systems with infinite-dimensional Hilbert state spaces and its detection via
entanglement witnesses based on second moments.

\section{Continuous-variable entanglement}\label{sec_CV_entanglement}

We assume that the considered systems $1,2,\ldots,N$ are the CV systems. Each of the systems is then
fully characterized by the position and momentum operators $x_j$ and $p_j$ ($[x_j,p_k]=i\delta_{j,k}$), $j=1,2,\ldots,N$,
respectively. From now on, we work with the optical platform, where the systems are realized by modes and
the position and momentum operators by the respective quadrature operators. It is further convenient
to introduce the column vector $\xi=\left(\xi_{x}^{T},\xi_{p}^{T}\right)^T$, where $\xi_{x}=\left(x_{1},x_{2},\dots,x_N\right)^T$ and
$\xi_{p}=\left(p_{1},p_{2},\dots,p_N\right)^T$, which allows us to express
the quadrature commutation rules in the compact form $[\xi_j,\xi_k]=i(\Omega_{N})_{jk}$ with
\begin{equation}\label{OmegaN}
\Omega_{N}=\begin{pmatrix}\mathbb{O}_N & \openone_N\\ -\openone_N & \mathbb{O}_N\end{pmatrix},
\end{equation}
where $\openone_N$ and $\mathbb{O}_N$ is the $N\times N$ identity matrix and zero matrix, respectively.
A particularly important class of states of CV systems is given by the so-called Gaussian states,
which are defined as states with a Gaussian-shaped Wigner phase-space distribution. For $N$ modes the states are
therefore fully described by the $2N\times1$ vector $\langle\xi\rangle=\mathrm{Tr}(\xi\rho)$ of the first
moments and by the $2N\times2N$ real symmetric covariance matrix (CM) $\gamma$ with entries
$(\gamma)_{jk}=\langle\{\Delta \xi_{j},\Delta \xi_k\}\rangle$, where $\{A,B\}\equiv AB+BA$ is the anticommutator
and $\Delta A\equiv A-\langle A\rangle$. From the definition of the CM it further follows that it
obeys the uncertainty principle \cite{Simon_00}
\begin{equation}\label{ineq:Heisenbergg}
\gamma+i\Omega_{N}\geq0.
\end{equation}

The CM contains information on correlations of the corresponding state and can be used to detect
GME by means of the biseparability criterion \cite{Hyllus_06}. It says that if an $N$-mode
state with CM $\gamma_{\rm BS}$ is biseparable, then there exist bipartitions $I_{k}|J_{k}$
and CMs $\gamma_{I_{k}}\oplus\gamma_{J_{k}}$ which are block diagonal with respect to the bipartition
$I_{k}|J_{k}$, and probabilities $\lambda_{k}$ such that
\begin{equation}\label{CMbiseparability}
\gamma_{\rm BS}-\sum_{k=1}^{K_N}\lambda_{k}\left(\gamma_{I_{k}}\oplus\gamma_{J_{k}}\right)\geq0.
\end{equation}
If the criterion (\ref{CMbiseparability})
is violated, the corresponding state is GME. An example of the Gaussian GME state is provided by the
CV GHZ-like state \cite{vanLoock_00} with CM
\begin{equation}\label{GHZ}
\gamma^{\rm GHZ}=\begin{pmatrix}
a_{+} & c_{+} & c_{+}\\
c_{+} & a_{+} & c_{+}\\
c_{+} & c_{+} & a_{+}
\end{pmatrix}\oplus\begin{pmatrix}
a_{-} & c_{-} & c_{-}\\
c_{-} & a_{-} & c_{-}\\
c_{-} & c_{-} & a_{-}
\end{pmatrix},
\end{equation}
where
\begin{equation}\label{ac}
a_{\pm}=\frac{e^{\pm 2r}+2e^{\mp 2r}}{3},\quad c_{\pm}=\frac{e^{\pm 2r}-e^{\mp 2r}}{3}
\end{equation}
and $r\geq0$ is the squeezing parameter. The GHZ state with CM (\ref{GHZ}) is also fully inseparable,
because the state is pure and for pure states the concepts of GME and full inseparability coincide.

The situation changes for mixed states, where similar to qubits  \cite{Piani_07} one can find also CV
fully inseparable states which are biseparable. One example is given
by the manifestly biseparable state $\left(\rho_{12}^{\rm TMSV}\otimes\rho_{3}^{\rm sq}+\rho_{1}^{\rm sq}\otimes\rho_{23}^{\rm TMSV}\right)/2$
\cite{Shalm_13,Teh_14}. Here, $\rho_{i}^{\rm sq}$ is the single-mode squeezed state of mode $i$ with the squeezing parameter $r$ and
$\rho_{jk}^{\rm TMSV}$ is the two-mode squeezed vacuum (TMSV) state of modes $j$ and $k$ with CM
\begin{equation}\label{TMSV}
\gamma_{jk}^{\rm TMSV}=\begin{pmatrix}
a & c \\
c & a
\end{pmatrix}\oplus\begin{pmatrix}
a & -c \\
-c & a
\end{pmatrix},
\end{equation}
where $a=\cosh(2r)$ and $c=\sinh(2r)$. An even simpler fully symmetrical example is provided by the state
\begin{align}\label{eq:rhotestBisep}
\rho^{\rm test}=&\frac{1}{3}\left(\rho^{\rm TMSV}_{12}\otimes|0\rangle_{3}\langle0|+\rho_{13}^{\rm TMSV}\otimes|0\rangle_{2}\langle0|\right.\nonumber\\
&\left.+|0\rangle_{1}\langle0|\otimes\rho_{23}^{\rm TMSV}\right),
\end{align}
where all TMSV states have the same squeezing parameter $r$. Again, the state is obviously biseparable.
Moreover, by applying the positive partial transposition criterion \cite{Simon_00} on CM of the state,
\begin{equation}\label{eq:testBisep}
\gamma^{\rm test}=\frac{1}{3}(\gamma^{\rm TMSV}_{12}\oplus\openone_3+\gamma_{13}^{\rm TMSV}\oplus\openone_{2}+\openone_{1}\oplus\gamma_{23}^{\rm TMSV}),
\end{equation}
one can show (see Appendix~\ref{appendix_sec_I} for details of the calculation) that for $r\in(0,1.24)$ the state (\ref{eq:rhotestBisep})
is at the same time entangled across all three bipartite splits $1|23, 2|31$ and $3|12$,
and thus it is fully inseparable as required.

Both the GHZ-like state with CM (\ref{GHZ}) as well as the state with CM (\ref{eq:rhotestBisep}) are simple states suitable for
testing of the CV GME criteria. In the following sections we develop such criteria based
only on variances of two-mode quadrature combinations and test them, among other things, on these states.

\section{Minimal criteria for genuine multipartite entanglement}\label{sec_minimal_criteria}

We start by recalling that we look for the GME criteria based on second moments of
quadrature operators. Additionally, we restrict ourselves for a while to the sum criteria,
for which the left-hand side (LHS) is linear in the second moments. Introducing $M$ quadrature linear
combinations
\begin{equation}\label{wi}
w_{i}=\sum_{j=1}^{2N}R_{ji}\xi_{j},
\end{equation}
$i=1,2,\ldots,M$, where $R_{ji}$ are elements of a real $2N\times M$ matrix, the LHS of the criterion
then can be expressed for the state with CM $\gamma$ in the form
\begin{equation}\label{sumw}
\sum_{i=1}^{M}\langle\left(\Delta w_{i}\right)^{2}\rangle=\frac{1}{2}\mbox{Tr}\left[\gamma RR^{T}\right],
\end{equation}
where the symbol $\mbox{Tr}$ stands for the matrix trace. Further, making use of the structure of biseparable states
(\ref{eq:rhobisep}) together with the Cauchy-Schwarz inequality, we can arrive at the following generic form of the
necessary condition for biseparability:
\begin{equation}\label{Zcriterion}
\mbox{Tr}[\gamma_{BS}Z]\geq c,
\end{equation}
for all CMs $\gamma_{BS}$ of biseparable states. Here, $Z=RR^{T}$ is a real, symmetric $2N\times 2N$ matrix
satisfying the condition $Z\geq0$ and $c$ is a non-negative constant which depends on matrix $R$. Provided that
the matrix $Z$ has suitable properties and the inequality (\ref{Zcriterion}) is violated by some physical CM,
the matrix $Z/c$ can be interpreted as a GME witness in the space of second moments \cite{Anders_03,Hyllus_06}.
Expression of the inseparability criterion in terms of the witness matrix is advantageous as it allows
us to impose consistently further constraints on its structure, e.g., that the LHS of the criterion
contains only the minimal number of two-mode quadrature combinations. In the next subsection we derive such
the criteria. The derivation consists of two parts. In the first part, we find the structure of the LHS,
whereas in the second part we derive the right-hand side (RHS) of the criterion, which is given by the
lower bound on the LHS for all biseparable states.

\subsection{Left-hand side of the criterion}

At the outset we derive the LHS of the sought GME criterion. Similar to other entanglement criteria we
rule out combinations mixing position and momentum quadratures by assuming $Z=Z^{x}\oplus Z^{p}$, where
$Z^{x}$ and $Z^{p}$ are real, symmetric, and positive-semidefinite $N\times N$ matrices.
Inspired by the position part of the criterion (\ref{Shalm}), which involves
combinations $x_{1}-x_{2}$ and $x_{2}-x_{3}$, we further seek the criteria containing the least number of only
two-mode quadrature combinations. We therefore require the criteria to utilize only the minimal number of two-mode
reduced CMs (marginal CMs), which are needed for detection of GME. The marginal CMs comprise the so-called
minimal set of marginals and its structure is already known \cite{Paraschiv_17,Nordgren_22}. The set has to contain all
modes and one cannot divide it into a subset and its complement without having a common mode.

A more instructive graphical representation of the minimal set is obtained using the tools of graph theory \cite{West_01}.
Recall, that an undirected graph of order $N$ is a pair $G=(V,E)$ of a
set $V=\{1,2,\ldots,N\}$ of $N$ vertices, and a set $E\subseteq K\equiv\{\{u,v\}|(u,v)\in V^2 \wedge u\ne v\}$ of edges.
In our case a vertex $j$ of the graph represents mode $j$, whereas the edge connecting adjacent vertices $j$ and $k$
represents marginal CM $\gamma_{jk}$ of modes $j$ and $k$. The minimal set of marginals is then represented by
an undirected connected graph containing no cycles, also know as an unlabeled tree \cite{Steinbach_99},
which shall be henceforth denoted as $T$. Here, we are interested in the GME which exists in systems of at least
three modes and we therefore assume $N\geq3$ from now. A closed formula for the number of non-isomorphic trees of order
$N$ is not known, yet it is known to grow exponentially \cite{Otter_48}. For small $N$ the number can be found in
Ref.~\cite{A000055}, and in particular, for $N=3,4,5,6,\ldots,$ there is $1, 2, 3, 6, \ldots,$ non-isomorphic unlabeled trees.

In the next step we imprint the structure of the respective minimal set of marginals on the LHS of the required criterion.
This can be done elegantly with the help of the adjacency matrix of a graph $G$, denoted as $A$, which is an $N\times N$
symmetric matrix in which entry $A_{jk}$ is the number of edges in $G$ connecting vertices $j$ and $k$. Since a tree is a
simple graph with no loops or multiple edges, its adjacency matrix is a zero-one matrix with zeros on the diagonal \cite{West_01}.
In order the criterion to use only the minimal set of marginals corresponding to a tree with adjacency matrix $A$, its matrices
denoted as $Z_{A}^{x}$ and $Z_{A}^{p}$ have to have generally non-zero diagonal elements and zero off-diagonal elements on the
same places as the matrix $A$. This is simply achieved by the matrices
\begin{equation}\label{ZAxp}
Z_{A}^{x,p}=(\openone_{N}+A)\circ Z^{x,p},
\end{equation}
where $X\circ Y$ is the Hadamard (elementwise) matrix product with entries $(X\circ Y)_{jk}=X_{jk}Y_{jk}$ \cite{Halmos_48} and $Z^{x,p}$ are some real symmetric
matrices with no zero entries.
The LHS of the required criterion corresponding to a minimal set of marginal CMs with
adjacency matrix $A$ thus possesses the following structure:
\begin{equation}\label{LHS1}
\mbox{Tr}[\gamma Z_{A}]=\mbox{Tr}[\gamma^{x}Z_{A}^{x}]+\mbox{Tr}[\gamma^{p}Z_{A}^{p}].
\end{equation}
Here $Z_{A}^{x,p}$ are some real, symmetric, and positive-semidefinite matrices  of
the form (\ref{ZAxp}), and $\gamma^{x}$ and $\gamma^{p}$ is the first and second $N\times N$ diagonal block of the
CM $\gamma$ with entries $(\gamma^{x})_{jk}=\langle\{\Delta x_{j},\Delta x_{k}\}\rangle$ and
$(\gamma^{p})_{jk}=\langle\{\Delta p_{j},\Delta p_{k}\}\rangle$, respectively.

It remains to rewrite the RHS of Eq.~(\ref{LHS1}) in terms of variances of two-mode quadrature combinations.
Since the witness matrices $Z_{A}^{x,p}$ are always real, symmetric and positive-semidefinite, one can use
the Cholesky decomposition and express them as $Z_{A}^{\alpha}=L_{A}^{\alpha}(L_{A}^{\alpha})^{T}$, where $L_{A}^{\alpha}$, $\alpha=x,p$, are real
lower triangular $N\times N$ matrices with non-negative diagonal elements \cite{Horn_85,Golub_96,Dongarra_79}.
Like in the case of the matrix $Z_{A}^{\alpha}$ the structure of the Cholesky matrix $L_{A}^{\alpha}$ also depends on the respective
adjacency matrix $A$, which is expressed by the lower index $A$. In what follows, the lower index is dropped for brevity, i.e., from now
by the symbol $L^{\alpha}$ we understand a Cholesky matrix corresponding to some adjacency matrix $A$.

Making further use of the cyclic property of the matrix trace and the definition of a CM we then get
\begin{align}\label{eq:trace}
\mathrm{Tr}[\gamma^{\alpha}Z_{A}^{\alpha}]=&\mathrm{Tr}[\gamma^{\alpha}L^{\alpha}(L^{\alpha})^T]\nonumber\\
=&2\sum_{i=1}^{N}\langle\{\Delta[(L^{\alpha})^T\xi_{\alpha}]_i\}^2\rangle\nonumber\\
=&2\sum_{i=1}^{N}\langle\left(\Delta u_i^{\alpha}\right)^2\rangle\equiv2U^{\alpha},
\end{align}
$\alpha=x,p$. Here we introduced the multi-mode position and momentum variables,
\begin{equation}\label{uxpi}
u_{i}^{x}\equiv\sum_{j=i}^{N}L^{x}_{ji}x_{j},\quad u_{i}^{p}\equiv\sum_{j=i}^{N}L^{p}_{ji}p_{j},
\end{equation}
$i=1,\ldots,N$, and sums of their variances,
\begin{equation}\label{Uxp}
U^{x}\equiv\sum_{i=1}^{N}\langle\left(\Delta u_i^{x}\right)^2\rangle,\quad U^{p}\equiv\sum_{i=1}^{N}\langle\left(\Delta u_i^{p}\right)^2\rangle.
\end{equation}
Similar to other entanglement criteria we then define the LHS of our criterion as one half of the expression (\ref{LHS1}), i.e., $U^{x}+U^{p}$.
However, in the present case each of the quantities $U^{\alpha}$, $\alpha=x,p$, contains $N$ variances of the quadrature combinations of the form (\ref{uxpi}).

In general, the LHS will not still possess the desired form containing variances of at most two-mode quadrature combinations. Namely, for $i=1,2,\ldots,N-2$
the operators (\ref{uxpi}) contain generally $N-i+1 \geq 2$ terms and not at most two, as required. Interestingly, also all these
combinations can attain the required structure, at least for some trees, if we label suitably vertices of the tree representing
the considered minimal set of marginal CMs. The labeling can be found in several steps.

At the outset we show how the requirement that quadrature combinations (\ref{uxpi}) contain at most two terms
manifests itself on the level of the respective adjacency matrix $A$. Recall first, that we seek a necessary
condition for biseparability of the form $f_{\gamma}(L^{x},L^{p})\geq g(L^{x},L^{p})$.
On the LHS there is a function $f_{\gamma}(L^{x},L^{p})$ which depends on CM $\gamma$ of the tested state
and Cholesky matrices $L^{x,p}$ of some witness matrices $Z_{A}^{x,p}$. The RHS of the condition, which we derive in the next section, imposes
a lower bound on the LHS for all biseparable states, and it is formed by a function $g(L^{x},L^{p})$ depending only on the matrices $L^{x,p}$.
The biseparability condition then can be used for testing of the presence of GME in a given CM $\gamma$ by numerical minimization
of the difference $f_{\gamma}(L^{x},L^{p})-g(L^{x},L^{p})$ over the matrices $L^{x,p}$. If the minimum is negative, i.e.,
the biseparability condition is violated, then the state with CM $\gamma$ is inevitably GME. Here, we are interested only in
a generic form of the biseparability condition capable of detecting the largest number of GME states from a minimal set of marginal CMs. Clearly, such a condition is
obtained if the function $f_{\gamma}(L^{x},L^{p})$ depends on the maximum number of independent variables, i.e., if the Cholesky matrices $L^{x,p}$
corresponding to the minimal set have the maximum number of non-zero elements. Working with the positive-semidefinite matrices $Z_{A}^{x,p}$,
we would have to take into account that some diagonal elements of the Cholesky matrices $L^{x,p}$ can be equal to zero
\cite[p.~8.3]{Dongarra_79} thereby reducing the number of the variables over which we will optimize. For this reason we assume from now that the
matrices $Z_{A}^{x,p}$ are positive-definite which implies that the corresponding Cholesky matrices $L^{x,p}$ possess strictly positive diagonal elements which can be expressed as
\cite[Theorem~4.2.5]{Golub_96}
\begin{equation}\label{Lii}
L^{\alpha}_{11}=\sqrt{(Z_{A}^{\alpha})_{11}},\quad L^{\alpha}_{ii}=\sqrt{(Z_{A}^{\alpha})_{ii}-\sum_{j=1}^{i-1}(L_{ij}^{\alpha})^2},
\end{equation}
$i=2,3,\ldots,N$.

Now, let us rewrite the first $N-1$ combinations (\ref{uxpi}) in the compact form
\begin{align}\label{uxpi2}
u_{i}^{x}=&x_{i}L^{x}_{ii}+(x_{i+1},x_{i+2},\ldots,x_{N})\cdot L^{x}_{i+1:N,i},\nonumber\\
u_{i}^{p}=&p_{i}L^{p}_{ii}+(p_{i+1},p_{i+2},\ldots,p_{N})\cdot L^{p}_{i+1:N,i},
\end{align}
$i=1,2,\ldots,N-1$, where the symbol ``$\,\cdot\,$'' stands for the scalar product and
\begin{equation}\label{Livect}
L^{\alpha}_{i+1:N,i}\equiv(L^{\alpha}_{i+1i},L^{\alpha}_{i+2i},\ldots,L^{\alpha}_{Ni})^{T}
\end{equation}
are $(N-i)\times 1$ column vectors comprising lower triangular part without the main diagonal of the matrix $L^{\alpha}$.
Hence, we see that the operators (\ref{uxpi2}) will contain at most two non-zero terms if each of the vectors $L^{\alpha}_{i+1:N,i}$ will have
at most one non-zero component. In the Appendix~\ref{appendix_sec_new} we show that this will be the case if all vectors $(Z_{A}^{\alpha})_{i+1:N,i}$, $i=1,2,\ldots,N-1$ will possess at most one non-zero component and the vector (\ref{Livect}) is the given by
\begin{equation}\label{Li2}
L^{\alpha}_{i+1:N,i}=\frac{(Z_{A}^{\alpha})_{i+1:N,i}}{L_{ii}^{\alpha}}.
\end{equation}
Making further use of the formula (\ref{ZAxp}) we can write
\begin{equation}\label{Zi}
(Z_{A}^{\alpha})_{i+1:N,i}=A_{i+1:N,i}\circ(Z^{\alpha})_{i+1:N,i},
\end{equation}
which leads, when combined with Eq.~(\ref{Li2}), to the final formula for the Cholesky matrices of the sought criterion,
\begin{equation}\label{LAalpha}
L^{\alpha}=(\openone_{N}+A)\circ\LL^{\alpha},
\end{equation}
where $\LL^{\alpha}$ is a lower triangular matrix with non-zero entries and strictly positive diagonal elements. Ascertaining of the structure of the adjacency matrix
$A$, which leads to the quadrature combinations (\ref{uxpi}) with at most two non-zero terms, is now simple. The quadrature combinations (\ref{uxpi}) then contain only $4N-2$ elements.
From the formula (\ref{Zi}) it follows immediately that this happens if the vectors $A_{i+1:N,i}$, $i=1,2,\ldots,N-1$,
contain only one non-zero element. As the structure of the adjacency matrix is for a given tree dictated by the labeling of its
vertices, the question of the possibility to have a GME criterion based only on at most two-mode quadrature combinations boils down
to the question of the existence of the tree labeling which would possess in each vector $A_{i+1:N,i}$ only one
non-zero component. In what follows we show that such a labeling really exists, at least for some trees.

\subsubsection{Linear tree}

\begin{figure}[h]
\begin{tikzpicture}[node distance   = 0.25 cm]
$a)$\hspace{0.8cm}
    \tikzset{VertexStyle/.style = {shape          = circle,
                                    thick,
                                    draw            = black,
                                    text           = black,
                                    inner sep      = 0.25pt,
                                    outer sep      = 0pt,
                                    minimum size   = 11 pt}}
    \tikzset{EdgeStyle/.style   = {thin,
                                    double          = black,
                                    double distance = 0.125pt}}
        \node[](a){};
        \node[,right=of a](b){};
        \node[,right=of b](c){};
        \node[,right=of c](d){};
        \node[VertexStyle,below= 0.1 cm of a](A){\footnotesize \textbf{1}};
        \node[VertexStyle,right=of A](B){\footnotesize \textbf{2}};
        \node[VertexStyle,right=of B](C){\footnotesize \textbf{3}};
        \node[VertexStyle,right=of C](D){\footnotesize \textbf{4}};
        \draw[EdgeStyle](C) to node{ } (D);
        \draw[EdgeStyle](A) to node{ } (B);
        \draw[EdgeStyle](B) to node{ } (A);
        \draw[EdgeStyle](B) to node{ } (C);
\hspace{3.5cm} $b)$ \hspace{0.8cm}
        \node[](a){};
        \node[,right=of a](b){};
        \node[,right=of b](c){};
        \node[,right=of c](d){};
        \node[VertexStyle,below= 0.1 cm of a](A){\footnotesize \textbf{3}};
        \node[VertexStyle,right=of A](B){\footnotesize \textbf{4}};
        \node[VertexStyle,right=of B](C){\footnotesize \textbf{2}};
        \node[VertexStyle,right=of C](D){\footnotesize \textbf{1}};
        \draw[EdgeStyle](C) to node{ } (D);
        \draw[EdgeStyle](A) to node{ } (B);
        \draw[EdgeStyle](B) to node{ } (A);
        \draw[EdgeStyle](B) to node{ } (C);
\end{tikzpicture}
\caption{Four-mode example of the standard labeling $a)$ and the alternative reverse level order labeling $b)$ of a linear tree, which yields the quantities (\ref{Uxp})
with variances of quadrature combinations (\ref{uxpi}) containing at most two terms.}
\label{fig:lin_alternative}
\end{figure}
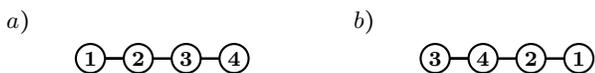

The first example of a tree for which the labeling is easy to find is an $N$-vertex linear tree. Consider the standard labeling of the
vertices by natural numbers in an increasing order from the leftmost vertex to the rightmost vertex as exemplified in Fig.~\ref{fig:lin_alternative}~a)
(see also labeled trees in the second, fourth and seventh row of Tab.~\ref{table0} for examples corresponding to $N=4, 5$ and $6$).
The corresponding adjacency matrix has elements $A_{ij}=\delta_{|i-j|,1}$, $i,j=1,2,\ldots,N$, for which the quantities (\ref{Uxp}) are given by
\begin{align}\label{Ualphalin}
U^{\alpha}_{\rm lin}=&\sum_{i=1}^{N-1}\langle[\Delta(\LL^{\alpha}_{ii}\alpha_i+\LL^\alpha_{i+1i}\alpha_{i+1})]^2\rangle\nonumber\\
&+(\LL^\alpha_{NN})^2\langle(\Delta \alpha_N)^2\rangle,
\end{align}
$\alpha=x,p$, where the relations (\ref{uxpi}), (\ref{Uxp}) and (\ref{LAalpha}) have been used. Clearly, the RHS of the latter formula consists of
variances of at most two-mode quadrature combinations as required. Note further, that the presented labeling is not the only one for which the
LHS of the criterion attains the desired form. For instance, it is easy to show that for the so-called reverse level order labeling introduced below in this paper
and depicted for a linear tree in Fig.~\ref{fig:lin_alternative}~b), the sums (\ref{Uxp}) also contain variances of at most two-mode quadrature combinations.
\subsubsection{Star tree}

Let us now move to the $N$-vertex star tree. We see that the labeling we are looking for is established if we label the central vertex with $N$
and the other vertices are labeled by the natural numbers $N-1,N-2,\ldots,1$ in a decreasing order in the anticlockwise direction
(see labeled trees in the third, fifth and eighth row of Tab.~\ref{table0} for examples corresponding to $N=4,5$ and $6$).
For this labeling the only non-zero elements of the adjacency matrix are $A_{Ni}=A_{iN}=1$, and they yield
sums (\ref{Uxp}) of the following form:
\begin{align}\label{Ualphastar}
U^{\alpha}_{\rm star}=&\sum_{i=1}^{N-1}\langle[\Delta(\LL^{\alpha}_{ii}\alpha_{i}+\LL^{\alpha}_{Ni}\alpha_{N})]^2\rangle\nonumber\\
&+(\LL^{\alpha}_{NN})^2\langle(\Delta \alpha_N)^2\rangle,
\end{align}
$\alpha=x,p$. Again, the LHS of the corresponding criterion comprise variances of at most two-mode quadrature combinations.

\subsubsection{Generic tree}

Interestingly, even generic trees can be labeled such, at least for low orders, i.e., low $N$, that the corresponding quadrature
combinations (\ref{uxpi}) contain at most two non-zero terms. This can be done as follows.

Let us start by recalling that the degree of a vertex in a graph is the number of edges incident to it and that a vertex of degree
one is called as a leaf. Consider now a tree $T$ and imagine an iterative procedure \cite{Hedetniemi_81,West_01} in which in each step
we delete all leaves of the tree obtained in the previous step. At the end, we are left either with a single vertex or a single edge,
which is the so-called center of the tree $T$ (see Tab.~\ref{tab:tree_examples} for examples of both the instances).
\begin{table}[]
    \caption{Examples of the reverse level order labeling for generic trees. The trees are rooted by selecting a central vertex or one endpoint of the central edge as a root (red vertices).
    The vertices are then labeled by natural numbers in a decreasing order staring by labeling the root with $N$ and then going from level to level, where in each
    level the vertices are labeled consecutively from left to right.}
    \centering
    \scalebox{0.75}{
    \begin{tabular}{|@{}c@{}|@{}c@{}|@{}c@{}|@{}c@{}|}
        \hline
        ~$N$~ & Unlabeled tree & ~Type of center~ & Rooted tree \\ \hline
        &  &  &  \\[-0.9em]
        \centered{$7$} & \centered{\begin{adjustbox}{valign=c}
        \begin{tikzpicture}[node distance   = 0.25 cm]
            \centering
          \tikzset{VertexStyle/.style = {shape          = circle,
                                        thick,
                                        draw            = black,
                                         text           = black,
                                         inner sep      = 0.25pt,
                                         outer sep      = 0pt,
                                         minimum size   = 11 pt}}
          \tikzset{EdgeStyle/.style   = {thin,
                                         double          = black,
                                         double distance = 0.125pt}}
             \node[](A){ };
             \node[VertexStyle,right=of A](B){ };
             \node[VertexStyle,right=of B](I){ };
             \node[VertexStyle,right=of I](C){ };
             \node[,right=of C](D){ };
             \node[VertexStyle,below= 0.1 cm of A](E){ };
             \node[VertexStyle,above= 0.1 cm of A](F){ };
             \node[VertexStyle,below= 0.1 cm of D](G){ };
             \node[VertexStyle,above= 0.1 cm of D](H){ };
             \draw[EdgeStyle](B) to node{ } (E);
             \draw[EdgeStyle](B) to node{ } (F);
             \draw[EdgeStyle](C) to node{ } (G);
             \draw[EdgeStyle](C) to node{ } (H);
             \draw[EdgeStyle](B) to node{ } (I);
             \draw[EdgeStyle](I) to node{ } (C);

          \end{tikzpicture}
        \end{adjustbox}} & \centered{vertex} & \centered{\begin{tikzpicture}[
                every node/.style = {shape = circle, thick, draw = black, text = black, inner sep = 0.25pt, outer sep = 0pt, minimum size = 11pt,font=\bfseries},
                level distance=0.7cm,
                level 1/.style={sibling distance=1.7cm},
                level 2/.style={sibling distance=1cm}]
                \node[shape = circle, thick, draw = red, text = red, inner sep = 0.25pt, outer sep = 0pt, minimum size = 11pt,font=\bfseries] {7}
                    child { node {6}
                        child { node {4}}
                        child { node{3}} }
                    child { node {5}
                      child { node {2}}
                      child { node {1} } };
            \end{tikzpicture}} \\ \hline
            &  &  &  \\[-0.9em]
         \centered{$7$} & \centered{
             \begin{tikzpicture}
                 [node distance   = 0.25 cm]
                \centering
              \tikzset{VertexStyle/.style = {shape          = circle,
                                            thick,
                                            draw            = black,
                                             text           = black,
                                             inner sep      = 0.25pt,
                                             outer sep      = 0pt,
                                             minimum size   = 11 pt}}
              \tikzset{EdgeStyle/.style   = {thin,
                                             double          = black,
                                             double distance = 0.125pt}}
                    \node[VertexStyle](A){};
                    \node[VertexStyle,right=of A](B){};
                    \node[VertexStyle,above= 0.1 cm of B](C){};
                    \node[VertexStyle,below= 0.1 cm of B](D){};
                    \node[VertexStyle,right=of B](E){};
                    \node[,right=of E](F){};
                    \node[VertexStyle,above= 0.1 cm of F](G){};
                    \node[VertexStyle,below= 0.1 cm of F](H){};
                    \draw[EdgeStyle](A) to node{ } (B);
                    \draw[EdgeStyle](B) to node{ } (C);
                    \draw[EdgeStyle](B) to node{ } (D);
                    \draw[EdgeStyle](B) to node{ } (E);
                    \draw[EdgeStyle](E) to node{ } (G);
                    \draw[EdgeStyle](E) to node{ } (H);
             \end{tikzpicture}
        } & \centered{edge} & \centered{\begin{tikzpicture}[
                every node/.style = {shape = circle, thick, draw = black, text = black, inner sep = 0.25pt, outer sep = 0pt, minimum size = 11pt,font=\bfseries},
                level distance=0.7cm,
                level 1/.style={sibling distance=1cm},
                level 2/.style={sibling distance=1cm}]
                \node[shape = circle, thick, draw = red, text = red, inner sep = 0.25pt, outer sep = 0pt, minimum size = 11pt,font=\bfseries] {7}
                    child { node {6}
                        child { node {3}}
                        child { node{2}}
                        child { node{1}} }
                    child { node {5}}
                    child { node {4}};
            \end{tikzpicture}} \\ \hline
            &  &  &  \\[-0.9em]
        \centered{$8$} & \centered{\begin{adjustbox}{valign=c}
        \begin{tikzpicture}[node distance   = 0.25 cm]
            \centering
          \tikzset{VertexStyle/.style = {shape          = circle,
                                        thick,
                                        draw            = black,
                                         text           = black,
                                         inner sep      = 0.25pt,
                                         outer sep      = 0pt,
                                         minimum size   = 11 pt}}
          \tikzset{EdgeStyle/.style   = {thin,
                                         double          = black,
                                         double distance = 0.125pt}}
             \node[](A){ };
             \node[VertexStyle,right=of A](B){ };
             \node[VertexStyle,right=of B](I){ };
             \node[VertexStyle,right=of I](J){ };
             \node[VertexStyle,right=of J](C){ };
             \node[,right=of C](D){ };
             \node[VertexStyle,below= 0.1 cm of A](E){ };
             \node[VertexStyle,above= 0.1 cm of A](F){ };
             \node[VertexStyle,below= 0.1 cm of D](G){ };
             \node[VertexStyle,above= 0.1 cm of D](H){ };
             \draw[EdgeStyle](B) to node{ } (E);
             \draw[EdgeStyle](B) to node{ } (F);
             \draw[EdgeStyle](C) to node{ } (G);
             \draw[EdgeStyle](C) to node{ } (H);
             \draw[EdgeStyle](B) to node{ } (I);
             \draw[EdgeStyle](I) to node{ } (J);
             \draw[EdgeStyle](J) to node{ } (C);

          \end{tikzpicture}
        \end{adjustbox}} & \centered{edge} & \centered{\begin{tikzpicture}[
                every node/.style = {shape = circle, thick, draw = black, text = black, inner sep = 0.25pt, outer sep = 0pt, minimum size = 11pt,font=\bfseries},
                level distance=0.7cm,
                level 1/.style={sibling distance=1.7cm},
                level 2/.style={sibling distance=1cm}]
                \node[shape = circle, thick, draw = red, text = red, inner sep = 0.25pt, outer sep = 0pt, minimum size = 11pt,font=\bfseries] {8}
                    child { node {7}
                        child { node {5}
                            child { node{2}}
                            child { node{1}} } }
                    child { node {6}
                      child { node {4}}
                      child { node {3} } };
            \end{tikzpicture}} \\ \hline
    \end{tabular}}
    \label{tab:tree_examples}
\end{table}
Alternatively, we get the center by taking a vertex or an edge which lies in the middle of any longest path in the tree \cite{Hedetniemi_81}.
Next, if the center of the tree is a vertex, we choose it as the so-called root of the tree, whereas
if it is an edge, we choose one of its endpoints as a root (see red vertices in Tab.~\ref{tab:tree_examples}). Having established a root, we
now associate with each vertex a level which is the number of edges from the vertex to the root and arrange the vertices into
horizontal layers, where in the topmost layer there is only the root, the second layer is comprised of level-one vertices etc.
(see the last column of Tab.~\ref{tab:tree_examples}). This allows us to introduce a labeling of the tree $T$, which can be called
fittingly as a reverse level order labeling. Here, we label the vertices by natural numbers in a decreasing order starting
by assigning the label $N$ to the root and continuing level by level going from left to right at each level (see the last column of Tab.~\ref{tab:tree_examples},
as well as the third column of  Tab.~\ref{table0} for explicit examples).

We checked by a direct calculation that for all configurations of up to $N=10$ the established reverse level order labeling leads to an adjacency matrix, where each vector $A_{i+1:N,i}$ contains only
one non-zero component and therefore the respective quadrature combinations (\ref{uxpi}) contain at most two non-zero terms. This leads us to a conjecture
that this statement holds generally for an arbitrary $N$-vertex tree.

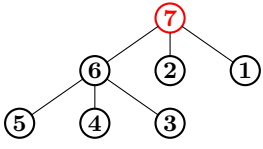
\begin{figure}[h]
\centering
\begin{tikzpicture}
                [every node/.style = {shape = circle, thick, draw = black, text = black, inner sep = 0.25pt, outer sep = 0pt, minimum size = 11pt,font=\bfseries},
                level distance=0.7cm,
                level 1/.style={sibling distance=1cm},
                level 2/.style={sibling distance=1cm}]
                \node[shape = circle, thick, draw = red, text = red, inner sep = 0.25pt, outer sep = 0pt, minimum size = 11pt,font=\bfseries] {7}
                    child { node{6}
                        child { node{5}}
                        child { node{4}}
                        child { node{3}} }
                    child { node{2}}
                    child { node{1}};
\end{tikzpicture}
\caption{Example of a tree labeling which is different from the reverse
level order labeling displayed in the second row of Tab.~\ref{tab:tree_examples}, yet it leads to the
GME criterion with at most two-mode quadrature combinations (\ref{uxpi}). See text for details.}
\label{fig:diflabeling}
\end{figure}

Note finally that the reverse level order labeling is not the only labeling yielding vectors $A_{i+1:N,i}$, $i=1,2,\ldots,N-1$,
each with only one non-zero component. Other labeling equipped with the same property is the ``natural'' labeling of a
linear tree exemplified in the preceding subsubsection or the labeling of a tree depicted in Fig.~\ref{fig:diflabeling}.

\subsection{Right-hand side of the criterion}

In the next step, we find a lower bound for the LHS of the sought criterion, which we obtained in the preceding subsection. Following the standard approach \cite{Teh_14},
we find a lower bound both on the sum $U^{x}+U^{p}$ and on the product $U^{x}U^{p}$ for all biseparable states. Thus we get two necessary conditions for
biseparability, which we call the sum condition and the product condition, respectively. Negation of the conditions then yields two
sufficient conditions for GME. Below we give examples of GME states detected by both the criteria, as well as by only one of them, which demonstrates that they certify GME of
intersecting but generally different sets of states.

In what follows we first derive a product necessary condition for biseparability and then we continue with derivation of the sum condition.

\subsubsection{Product criterion}

Our goal is to find a lower bound for the product
\begin{equation}\label{UxUp}
U^{x}U^{p}=\sum_{i,j=1}^{N}\langle\left(\Delta u_{i}^{x}\right)^2\rangle\langle\left(\Delta u_{j}^{p}\right)^2\rangle,
\end{equation}
where
\begin{equation}\label{ualpha}
u_{i}^{\alpha}=\sum_{j=1}^{N}L^{\alpha}_{ji}\alpha_{j},
\end{equation}
$\alpha=x,p$, for all biseparable states (\ref{eq:rhobisep}). Contrary to Eq.~(\ref{uxpi}), in Eq.~(\ref{ualpha}) we for simplicity do
not take into account explicitly the fact that $L_{ji}^{\alpha}=0$ for $j<i$.

Note first, that the product (\ref{UxUp}) is lower bounded as
\begin{align}\label{Uxpbisep}
U^{x}U^{p}&\stackrel{1}{\geq}\left(\sum_{k=1}^{K_{N}}\lambda_{k}U_{k}^{x}\right)\left(\sum_{l=1}^{K_{N}}\lambda_{l}U_{l}^{p}\right)\nonumber\\
&\stackrel{2}{\geq}\left(\sum_{k=1}^{K_N}\lambda_{k}\sqrt{U_{k}^{x}U_{k}^{p}}\right)^2,
\end{align}
with
\begin{equation}\label{Ukalpha}
U_{k}^{\alpha}\equiv\sum_{i=1}^{N}\langle\left(\Delta u_{i}^{\alpha}\right)^2\rangle_{\rho_{I_{k}|J_{k}}^{\rm sep}},
\end{equation}
$\alpha=x,p$, where the symbol $\langle\,\,\rangle_{\rho}$ stands for the mean in the state $\rho$.
Here, to get inequality $1$ we used the concavity of the variance with respect to convex mixtures \cite{Hofmann_03}, which we applied to the mixture $\rho^{\rm bisep}$, Eq.~(\ref{eq:rhobisep}),
whereas inequality $2$ follows from the Cauchy-Schwarz inequality.

Next, we lower bound the product
\begin{equation}\label{UkxUkp}
U_{k}^{x}U_{k}^{p}=\sum_{i,j=1}^{N}\langle\left(\Delta u_{i}^{x}\right)^2\rangle_{\rho_{I_{k}|J_{k}}^{\rm sep}}\langle\left(\Delta u_{j}^{p}\right)^2\rangle_{\rho_{I_{k}|J_{k}}^{\rm sep}}.
\end{equation}
For this purpose, consider a bipartite split $k\equiv I_{k}|J_{k}$, where $I_{k}=\{i_{1},i_{2},\ldots,i_{l}\}$ and $J_{k}=\{i_{l+1},i_{l+2},\ldots,i_{N}\}$, and
assume that the system is prepared in the separable state $\rho_{I_{k}|J_{k}}^{\rm sep}$, Eq.~(\ref{eq:rhosep}). Let as further decompose the operators
(\ref{ualpha}) into two parts, one belonging to the subsystem $I_{k}$ and the other to the subsystem $J_{k}$,
\begin{equation}\label{ualphasplit}
u_{i}^{\alpha}=u_{i,I_{k}}^{\alpha}+u_{i,J_{k}}^{\alpha},
\end{equation}
where
\begin{equation}\label{ualphaIJ}
u_{i,I_{k}}^{\alpha}\equiv\sum_{n=1}^{l}L^{\alpha}_{i_{n}i}\alpha_{i_{n}},\quad u_{i,J_{k}}^{\alpha}\equiv\sum_{n=l+1}^{N}L^{\alpha}_{i_{n}i}\alpha_{i_{n}}.
\end{equation}
Inspired by the approach of Ref.~\cite{Armstrong_15} we can then lower bound the variance product appearing on the RHS of Eq.~(\ref{UkxUkp}) as
\begin{align}\label{ukxkpbound}
&\langle(\Delta u_{i}^{x})^2\rangle_{\rho_{I_{k}|J_{k}}^{\rm sep}}\langle(\Delta u_{j}^{p})^2\rangle_{\rho_{I_{k}|J_{k}}^{\rm sep}}\nonumber\\
&\stackrel{1}{\geq}\left[\sum_l p_l^{(k)}\langle\left(\Delta u_{i}^{x}\right)^2\rangle_{\rho_{I_{k}\otimes J_{k}}^{(l)}}\right]\left[\sum_m p_m^{(k)}\langle\left(\Delta u_{j}^{p}\right)^2\rangle_{\rho_{I_{k}\otimes J_{k}}^{(m)}}\right]\nonumber\\
&\stackrel{2}{\geq}\left[\sum_l p_l^{(k)}\sqrt{\langle\left(\Delta u_{i}^{x}\right)^2\rangle_{\rho_{I_{k}\otimes J_{k}}^{(l)}}\langle\left(\Delta u_{j}^{p}\right)^2\rangle_{\rho_{I_{k}\otimes J_{k}}^{(l)}}}\right]^2\nonumber\\
&\stackrel{3}{\geq}\left\{\sum_l p_l^{(k)}\left[\sqrt{\langle(\Delta u_{i,I_{k}}^{x})^2\rangle_{\rho_{I_{k}}^{(l)}}\langle(\Delta u_{j,I_{k}}^{p})^2\rangle_{\rho_{I_{k}}^{(l)}}}\right.\right.\nonumber\\
&\hspace{0.5cm}+\left.\left.\sqrt{\langle(\Delta u_{i,J_{k}}^{x})^2\rangle_{\rho_{J_{k}}^{(l)}}\langle(\Delta u_{j,J_{k}}^{p})^2\rangle_{\rho_{J_{k}}^{(l)}}}\right]\right\}^2\nonumber\\
&\stackrel{4}{\geq}\frac{1}{4}\left\{\left|\left[(L^{x})^{T}L^{p}\right]_{ij}^{(I_{k})}\right|+\left|\left[(L^{x})^{T}L^{p}\right]_{ij}^{(J_{k})}\right|\right\}^2\nonumber\\
&\stackrel{5}{=}\frac{1}{4}\left\{\left[(L^{x})^{T}L^{p}\right]_{ij}^{2}+2(|X_{ij}^{(k)}|-X_{ij}^{(k)})\right\},
\end{align}
where we introduced the denotation $\rho_{I_{k}\otimes J_{k}}^{(l)}\equiv\rho_{I_{k}}^{(l)}\otimes\rho_{J_{k}}^{(l)}$. Here, inequality $1$ follows from
concavity of the variance, inequality $2$ is a consequence of the Cauchy-Schwarz inequality and inequality $3$ results from the averaging in the product
state $\rho_{I_{k}\otimes J_{k}}^{(l)}$ and again the Cauchy-Schwarz inequality. Finally, in inequality $4$ we used the uncertainty relations and the quantities appearing on the RHS of the inequality and equality $5$ are defined as
\begin{align}
\left[(L^{x})^{T}L^{p}\right]_{ij}^{(I_{k})}&\equiv\sum_{n=1}^{l}\left[(L^{x})^{T}\right]_{ii_{n}}L^{p}_{i_{n}j},\label{prodIk}\\
\left[(L^{x})^{T}L^{p}\right]_{ij}^{(J_{k})}&\equiv\sum_{n=l+1}^{N}\left[(L^{x})^{T}\right]_{ii_{n}}L^{p}_{i_{n}j},\label{prodJk}\\
X_{ij}^{(k)}&\equiv\left[(L^{x})^{T}L^{p}\right]_{ij}^{(I_{k})}\left[(L^{x})^{T}L^{p}\right]_{ij}^{(J_{k})}.\nonumber\\
\label{Xk}
\end{align}

Combining now Eq.~(\ref{UkxUkp}) with inequality (\ref{ukxkpbound}) we get immediately the following inequality
\begin{equation}\label{UkxUkpbound}
U_{k}^{x}U_{k}^{p}\geq\frac{1}{4}\left(\mathcal{L}+\mathcal{L}^{(k)}\right),
\end{equation}
where
\begin{align}
\mathcal{L}&\equiv\sum_{i,j=1}^{N}\left[(L^{x})^{T}L^{p}\right]_{ij}^{2},\label{L}\\
\mathcal{L}^{(k)}&\equiv2\sum_{i,j=1}^{N}(|X_{ij}^{(k)}|-X_{ij}^{(k)}).\label{Lk}
\end{align}
Finally, putting inequalities (\ref{Uxpbisep}) and (\ref{UkxUkpbound}) together, we arrive at the sought
necessary condition for biseparability of the form:
\begin{equation}\label{criterion}
U^{x}U^{p}\geq\frac{1}{4}\left\{\mathcal{L}+\mbox{min}\left[\mathcal{L}^{(1)},\ldots,\mathcal{L}^{(2^{N-1}-1)}\right]\right\},
\end{equation}
where we used the fact that $K_{N}=2^{N-1}-1$.

\subsubsection{Sum criterion}

Derivation of the sum condition is simpler. First, we use concavity of the variance to get
\begin{equation}\label{UxUpsum}
U^{x}+U^{p}\geq\sum_{i=1}^{N}\sum_{k=1}^{K_{N}}\lambda_{k}\sum_{\alpha=x,p}\langle\left(\Delta u_{i}^{\alpha}\right)^2\rangle_{\rho_{I_{k}|J_{k}}^{\rm sep}}.
\end{equation}
Next, we lower bound the sum of the variances on the RHS as
\begin{align}\label{uxupsumbound}
&\sum_{\alpha=x,p}\langle\left(\Delta u_{i}^{\alpha}\right)^2\rangle_{\rho_{I_{k}|J_{k}}^{\rm sep}}&\nonumber\\
&\stackrel{1}{\geq}\sum_{\alpha=x,p}\sum_lp_l^{(k)}\langle\left(\Delta u_{i}^{\alpha}\right)^2\rangle_{\rho_{I_{k}\otimes J_{k}}^{(l)}}\nonumber\\
&\stackrel{2}{=}\sum_{\alpha=x,p}\sum_lp_l^{(k)}\left[\langle(\Delta u_{i,I_{k}}^{\alpha})^2\rangle_{\rho_{I_{k}}^{(l)}}+\langle(\Delta u_{i,J_{k}}^{\alpha})^2\rangle_{\rho_{J_{k}}^{(l)}}\right]\nonumber\\
&\stackrel{3}{\geq}\left|\left[(L^{x})^{T}L^{p}\right]_{ii}^{(I_{k})}\right|+\left|\left[(L^{x})^{T}L^{p}\right]_{ii}^{(J_{k})}\right|,
\end{align}
where inequality $1$ again follows from concavity of the variance, whereas in equality $2$ we used
the decomposition (\ref{ualphasplit}) and the fact that we average over the product state $\rho_{I_{k}\otimes J_{k}}^{(l)}$.
To get inequality $3$ we used the inequality between the arithmetic and geometric mean, the uncertainty relations, and the definitions (\ref{prodIk}) and (\ref{prodJk}). Finally, if we
combine inequalities (\ref{UxUpsum}) and (\ref{uxupsumbound}) we obtain the desired sum criterion in the following form:
\begin{equation}\label{sumcriterion}
U^{x}+U^{p}\geq\mathrm{min}\left[\mathcal{K}^{(1)},\ldots,\mathcal{K}^{(2^{N-1}-1)}\right],
\end{equation}
where
\begin{equation}\label{calKk}
\mathcal{K}^{(k)}\equiv\sum_{i=1}^{N}\left\{\left|\left[(L^{x})^{T}L^{p}\right]_{ii}^{(I_{k})}\right|+\left|\left[(L^{x})^{T}L^{p}\right]_{ii}^{(J_{k})}\right|\right\}.
\end{equation}
Testing of the presence of GME in a given quantum state by means of the criteria (\ref{criterion}) and (\ref{sumcriterion})
requires to find a set of parameters $\LL^{\alpha}_{ij}$, for which the respective inequalities (\ref{criterion}) and
(\ref{sumcriterion}) are violated. This can be done by minimization of the difference
\begin{equation}\label{D}
D_{P}\equiv U^{x}U^{p}-\frac{1}{4}\left\{\mathcal{L}+\mbox{min}\left[\mathcal{L}^{(1)},\ldots,\mathcal{L}^{(2^{N-1}-1)}\right]\right\}
\end{equation}
or
\begin{equation}\label{DS}
D_{S}\equiv U^{x}+U^{p}-\mathrm{min}\left[\mathcal{K}^{(1)},\ldots,\mathcal{K}^{(2^{N-1}-1)}\right]
\end{equation}
over the parameters $\LL^{\alpha}_{ij}$. Clearly, if the difference is negative, the investigated state is GME.

In the next Section we use this approach to verify GME in several multi-mode quantum states. This requires to
select a certain tree and express the respective differences (\ref{D}) and (\ref{DS}) in terms of the parameters
$\LL^{\alpha}_{ij}$ by means of the formula (\ref{LAalpha}). This can be rather cumbersome, in particular,
for trees with many vertices. For this reason, it is desirable to develop a simpler construction of the criteria,
which would ideally utilize only the structure of the considered tree. In the following subsection we present
such a construction for the sum criterion.

\subsection{Direct construction of the sum criterion}

Practical utility of the proposed criteria would be significantly increased if we would be able to
construct them directly from the underlying tree. While construction of the LHS for both the criteria is the same, construction of the
RHS is transparent only for the sum criterion. Below, we therefore construct the RHS for the sum criterion,
whereas the construction for the product criterion is deferred for further research.

\subsubsection{Construction of the left-hand side}

\begin{figure}[h]
\centering
\begin{tabular}{lr}
\multicolumn{2}{c}{$i<j$}\\[-2pt]
\hspace{2pt}$\LL_{ii}^\alpha$ & $\LL_{jj}^\alpha$\\[-7pt]
\multicolumn{2}{c}{$\LL_{ji}^\alpha$}\\
\multicolumn{2}{l}{\begin{tikzpicture}[node distance   = 1.8 cm]
        \centering
        \tikzset{VertexStyle/.style = {shape          = circle,
                                        thick,
                                        draw            = black,
                                        text           = black,
                                        inner sep      = 0.25pt,
                                        outer sep      = 0pt,
                                        minimum size   = 11 pt}}
        \tikzset{EdgeStyle/.style   = {thin,
                                        double          = black,
                                        double distance = 0.125pt}}
            \node[VertexStyle](A){\footnotesize \textbf{$i$}};
            \node[,right=of A](B){};
            \node[VertexStyle,right= of B](C){\footnotesize \textbf{$j$}};
            \draw[EdgeStyle](A) to node{ } (C);
    \end{tikzpicture}} \\[-3pt]
\hspace{2pt}$\alpha_i$ & $\alpha_j$ \\[-5pt]
\multicolumn{2}{c}{$u_i^\alpha=\LL_{ii}^\alpha \alpha_i + \LL_{ji}^\alpha \alpha_j$}
\end{tabular}
\caption{Construction of the two-mode quadrature combination $u_{i}^{\alpha}$ corresponding to the edge $e_{ij}=ij$ connecting vertices $i$ and $j$. See text for details.}
\label{fig_ijedge}
\end{figure}
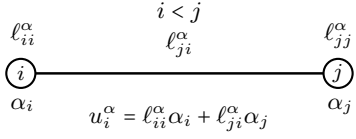

The quantities $U^{x,p}$, Eq.~(\ref{Uxp}), on the LHS of both criteria are the same and their construction is easy.
Namely, each $N$-vertex tree consists of $N-1$ edges \cite{West_01}.
Each of the edges connects a pair of vertices with labels $i$ and $j$, where by $i$ we
denote the label which is less than the label $j$, i.e., $i<j$. With each of the edges
we associate a two-mode quadrature linear combination $u_{i}^{\alpha}$ (see Fig.~\ref{fig_ijedge}).
Sum of the variances $\langle(\Delta u_{i}^{\alpha})^2\rangle$ over all edges then comprises
(up to the term $(\LL^{\alpha}_{NN})^2\langle(\Delta \alpha_N)^2\rangle$) the quantity $U^{\alpha}$.
More precisely, to each vertex $i$ we ascribe the quadrature $\alpha_{i}$, as well as the diagonal element $\LL_{ii}^{\alpha}$ of the matrix $\LL^{\alpha}$. Next, to each edge connecting
the vertices $i$ and $j$ we assign the off-diagonal entry $\LL_{ji}^{\alpha}$. The edge is then represented by the quadrature combination
\begin{equation}\label{ualphai}
u_{i}^{\alpha}=\LL^{\alpha}_{ii}\alpha_{i}+\LL^{\alpha}_{ji}\alpha_{j}.
\end{equation}
The quantity $U^{\alpha}$ in Eq.~(\ref{Uxp}) is then given by the sum of variances of all $N-1$ quantities (\ref{ualphai}),
plus the term $(\LL^{\alpha}_{NN})^2\langle(\Delta \alpha_N)^2\rangle$. Making use of this rule we can easily rederive the quantities
$U^{\alpha}$ for the linear tree and the star tree given in Eqs.~(\ref{Ualphalin}) and (\ref{Ualphastar}), respectively. In the next section we also use
the rule to construct the criteria corresponding to the more complex trees in the third column of  Tab.~\ref{table0}.

\subsubsection{Construction of the right-hand side}

Construction of the RHS of the sum criterion is also easy. Recall first, that in a tree with root $N$ there is for each vertex $i$ a unique $i,N$-path.
The so-called parent of vertex $i$ is an adjacent vertex on the $i,N$-path \cite{West_01} and each non-root vertex has exactly one parent. Let $j$ be the
label of the parent of vertex $i$, where inequality $i<j$ always holds in our labeling, and $e_{ij}\equiv ij$ be the edge joining the vertex $i$ with its parent
$j$. Now, for each non-root vertex $i$ we create a pair $(i,e_{ij})$ and represent it by the pair $(\LL^x_{ii}\LL^p_{ii},\LL^x_{ji}\LL^p_{ji})$
of products of elements of matrices $\LL^{x,p}$, which are associated with the vertex $i$ and the edge $e_{ij}$ according to Fig.~\ref{fig_ijedge}.
The root $N$ has no parent and thus there is no edge associated to it. Therefore, it is represented just by the term $\LL^x_{NN}\LL^p_{NN}$.

Moving to the construction of the RHS it is obvious that to every bipartite split $k\equiv I_{k}|J_{k}$ of $N$ modes into two disjoint subsets
$I_{k}=\{i_{1},i_{2},\ldots,i_{l}\}$ and $J_{k}=\{i_{l+1},i_{l+2},\ldots,i_{N}\}$ there corresponds exactly the same bipartite split of the set of all $N$ vertices
of the considered graph. The split divides the set of all $N-1$ pairs $(i,e_{ij})$ into two disjoint subsets, denoted as $\mathcal{U}$ and $\mathcal{C}$,
depending on whether the edge $e_{ij}$ is uncut by the split or it is cut off by the split. Formally, $(i,e_{ij})\in \mathcal{U}$ if $i,j\in I_{k}$
or $i,j\in J_{k}$, whereas $(i,e_{ij})\in \mathcal{C}$ if $i\in I_{k}$ and $j\in J_{k}$
or vice versa. Now, to each pair $(i,e_{ij})\in \mathcal{U}$ we ascribe the term $|\LL^x_{ii}\LL^p_{ii}+\LL^x_{ji}\LL^p_{ji}|$, whereas with each pair
$(i,e_{ij})\in \mathcal{C}$ we associate a different term $|\LL^x_{ii}\LL^p_{ii}|+|\LL^x_{ji}\LL^p_{ji}|$. The sum of all the latter terms for all
$N-1$ pairs $(i,e_{ij})$ plus the term $|\LL^x_{NN}\LL^p_{NN}|$, which corresponds to the root, gives finally the quantity $\mathcal{K}^{(k)}$, Eq.~(\ref{calKk}), appearing on the
RHS of the sum criterion (\ref{sumcriterion}).

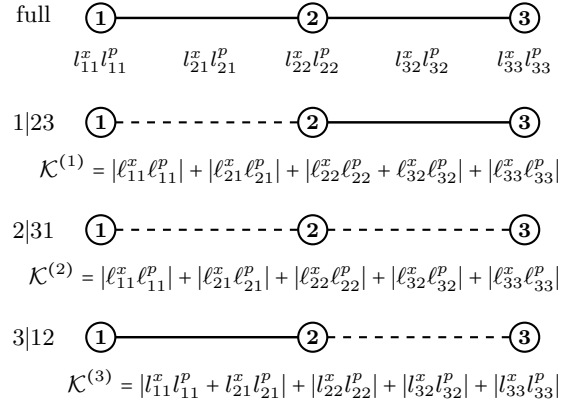
\begin{figure}[h]
\scalebox{0.93}{
\begin{tabular}{cc}
full & \multirow{2}{*}{\begin{tikzpicture}[node distance   = 1.1 cm]
        \centering
        \tikzset{VertexStyle/.style = {shape          = circle,
                                        thick,
                                        draw            = black,
                                        text           = black,
                                        inner sep      = 0.25pt,
                                        outer sep      = 0pt,
                                        minimum size   = 11 pt}}
        \tikzset{EdgeStyle1/.style   = {thin,
                                        double          = black,
                                        double distance = 0.125pt}}
        \tikzset{EdgeStyle2/.style  =  {thick,
                                        dashed,
                                        black}}
            \node[VertexStyle](A){\footnotesize \textbf{1}};
            \node[,right=of A](B){};
            \node[VertexStyle,right= of B](C){\footnotesize \textbf{2}};
            \node[, right=of C](D){};
            \node[VertexStyle, right=of D](E){\footnotesize \textbf{3}};
            \node[, below=0.1cm of A](a){\footnotesize $\LL^x_{11}\LL^p_{11}$};
            \node[, right=0.5cm of a](b){\footnotesize $\LL^x_{21}\LL^p_{21}$};
            \node[, below=0.1cm of C](c){\footnotesize $\LL^x_{22}\LL^p_{22}$};
            \node[, right=0.5cm of c](d){\footnotesize $\LL^x_{32}\LL^p_{32}$};
            \node[, below=0.1cm of E](e){\footnotesize $\LL^x_{33}\LL^p_{33}$};
            \draw[EdgeStyle1](A) to node{ } (C);
            \draw[EdgeStyle1](C) to node{ } (E);
    \end{tikzpicture}}\\
 & \\[20pt]
$1|23$ & \\[-15pt]
 & \multirow{2}{*}{\begin{tikzpicture}[node distance   = 1.1 cm]
        \centering
        \tikzset{VertexStyle/.style = {shape          = circle,
                                        thick,
                                        draw            = black,
                                        text           = black,
                                        inner sep      = 0.25pt,
                                        outer sep      = 0pt,
                                        minimum size   = 11 pt}}
        \tikzset{EdgeStyle1/.style   = {thin,
                                        double          = black,
                                        double distance = 0.125pt}}
        \tikzset{EdgeStyle2/.style  =  {thick,
                                        dashed,
                                        black}}
            \node[VertexStyle](A){\footnotesize \textbf{1}};
            \node[,right=of A](B){};
            \node[VertexStyle,right= of B](C){\footnotesize \textbf{2}};
            \node[, right=of C](D){};
            \node[VertexStyle, right=of D](E){\footnotesize \textbf{3}};
            \draw[EdgeStyle2](A) to node{ } (C);
            \draw[EdgeStyle1](C) to node{ } (E);
    \end{tikzpicture}}\\
  &  \\
\multicolumn{2}{r}{$\mathcal{K}^{(1)}=|\LL^x_{11}\LL^p_{11}|+|\LL^x_{21}\LL^p_{21}|+|\LL^x_{22}\LL^p_{22}+\LL^x_{32}\LL^p_{32}|+|\LL^x_{33}\LL^p_{33}|$} \\
 & \\[2pt]
$2|31$ & \\[-15pt]
 & \multirow{2}{*}{\begin{tikzpicture}[node distance   = 1.1 cm]
        \centering
        \tikzset{VertexStyle/.style = {shape          = circle,
                                        thick,
                                        draw            = black,
                                        text           = black,
                                        inner sep      = 0.25pt,
                                        outer sep      = 0pt,
                                        minimum size   = 11 pt}}
        \tikzset{EdgeStyle1/.style   = {thin,
                                        double          = black,
                                        double distance = 0.125pt}}
        \tikzset{EdgeStyle2/.style  =  {thick,
                                        dashed,
                                        black}}
            \node[VertexStyle](A){\footnotesize \textbf{1}};
            \node[,right=of A](B){};
            \node[VertexStyle,right= of B](C){\footnotesize \textbf{2}};
            \node[, right=of C](D){};
            \node[VertexStyle, right=of D](E){\footnotesize \textbf{3}};
            \draw[EdgeStyle2](A) to node{ } (C);
            \draw[EdgeStyle2](C) to node{ } (E);
    \end{tikzpicture}}\\
  &  \\
\multicolumn{2}{r}{$\mathcal{K}^{(2)}=|\LL^x_{11}\LL^p_{11}|+|\LL^x_{21}\LL^p_{21}|+|\LL^x_{22}\LL^p_{22}|+|\LL^x_{32}\LL^p_{32}|+|\LL^x_{33}\LL^p_{33}|$} \\
 & \\[2pt]
$3|12$ & \\[-15pt]
 & \multirow{2}{*}{\begin{tikzpicture}[node distance   = 1.1 cm]
        \centering
        \tikzset{VertexStyle/.style = {shape          = circle,
                                        thick,
                                        draw            = black,
                                        text           = black,
                                        inner sep      = 0.25pt,
                                        outer sep      = 0pt,
                                        minimum size   = 11 pt}}
        \tikzset{EdgeStyle1/.style   = {thin,
                                        double          = black,
                                        double distance = 0.125pt}}
        \tikzset{EdgeStyle2/.style  =  {thick,
                                        dashed,
                                        black}}
            \node[VertexStyle](A){\footnotesize \textbf{1}};
            \node[,right=of A](B){};
            \node[VertexStyle,right= of B](C){\footnotesize \textbf{2}};
            \node[, right=of C](D){};
            \node[VertexStyle, right=of D](E){\footnotesize \textbf{3}};
            \draw[EdgeStyle1](A) to node{ } (C);
            \draw[EdgeStyle2](C) to node{ } (E);
    \end{tikzpicture}}\\
 & \\
\multicolumn{2}{r}{$\mathcal{K}^{(3)}=|\LL^x_{11}\LL^p_{11}+\LL^x_{21}\LL^p_{21}|+|\LL^x_{22}\LL^p_{22}|+|\LL^x_{32}\LL^p_{32}|+|\LL^x_{33}\LL^p_{33}|$} \\
\end{tabular}}
\caption{Graphical construction of the quantities $\mathcal{K}^{(k)}$, $k=1,2,3$, appearing on the RHS of the sum criterion (\ref{sumcriterion})
for the $3$-vertex tree with the standard labeling (uppermost graph). The lower graphs correspond consecutively to the full tree after the cuts with respect
to bipartite splits $1\equiv1|23$, $2\equiv2|31$ and $3\equiv3|12$. The uncut edges are represented by solid lines, whereas the cut off edges
by the dashed lines. See text for details.}
\label{fig_RHSconstruction}
\end{figure}

Explicit construction of all the quantities $\mathcal{K}^{(k)}$ for the simplest case of $N=3$ modes is presented in Fig. \ref{fig_RHSconstruction}.

\section{Applications}\label{sec_applications}

\subsection{Three-mode states}

First, we test functionality of the product criterion (\ref{criterion}) and sum criterion (\ref{sumcriterion}) for the simplest case of $N=3$ modes.
Then there is only one linear tree which for the standard labeling yields the adjacency matrix given above Eq.~(\ref{Ualphalin}). Hence, one gets using
the formula (\ref{L})
\begin{align}\label{L3}
\mathcal{L}=&(\LL^x_{11}\LL^p_{11}+\LL^x_{21}\LL^p_{21})^2+(\LL^x_{21}\LL^p_{22})^2+(\LL^x_{22}\LL^p_{21})^2\nonumber\\
&+(\LL^x_{22}\LL^p_{22}+\LL^x_{32}\LL^p_{32})^2+(\LL^x_{32}\LL^p_{33})^2+(\LL^x_{33}\LL^p_{32})^2\nonumber\\
&+(\LL^x_{33}\LL^p_{33})^2.
\end{align}
The three modes can be divided into three bipartite splits which we denote as $1\equiv1|23$, $2\equiv2|31$ and $3\equiv3|12$,
and for the splits we further obtain
\begin{align}\label{Lk3}
\mathcal{L}^{(1)}=&2\left(|\LL^x_{11}\LL^p_{11}\LL^x_{21}\LL^p_{21}|-\LL^x_{11}\LL^p_{11}\LL^x_{21}\LL^p_{21}\right),\nonumber\\
\mathcal{L}^{(3)}=&2\left(|\LL^x_{22}\LL^p_{22}\LL^x_{32}\LL^p_{32}|-\LL^x_{22}\LL^p_{22}\LL^x_{32}\LL^p_{32}\right),\nonumber\\
\mathcal{L}^{(2)}=&\mathcal{L}^{(1)}+\mathcal{L}^{(3)},
\end{align}
by Eq.~(\ref{Lk}). Making use of the expressions for the quantities $U_{\rm lin}^{x,p}$, Eq.~(\ref{Ualphalin}), we get finally the minimal
three-mode product criterion for GME,
\begin{align}\label{pcriterion3}
\prod_{\alpha=x,p}&\left\{\sum_{i=1}^{2}\langle[\Delta(\LL^{\alpha}_{ii}\alpha_i+\LL^\alpha_{i+1i}\alpha_{i+1})]^2\rangle\right.\nonumber\\
&\quad\left.+\left(\LL^\alpha_{33}\right)^2\langle(\Delta \alpha_3)^2\rangle\right\}\nonumber\\
&\geq\frac{1}{4}\left\{\mathcal{L}+\mbox{min}\left[\mathcal{L}^{(1)},\mathcal{L}^{(2)},\mathcal{L}^{(3)}\right]\right\}.
\end{align}

Moving to the sum criterion (\ref{sumcriterion}) for three modes, the product on the LHS of the latter inequality is replaced with the sum,
whereas the quantities $\mathcal{K}^{(k)}$ appearing on the RHS were derived
in Fig.~\ref{fig_RHSconstruction}. Additionally, since $\mathcal{K}^{(1)},\mathcal{K}^{(3)}\leq \mathcal{K}^{(2)}$,
it is sufficient to take only $\mbox{min}\{\mathcal{K}^{(1)},\mathcal{K}^{(3)}\}$ on the RHS of the considered criterion. Putting it all together the
minimal three-mode sum criterion for GME reads as
\begin{align}\label{scriterion3}
\sum_{\alpha=x,p}&\left\{\sum_{i=1}^{2}\langle[\Delta(\LL^{\alpha}_{ii}\alpha_i+\LL^\alpha_{i+1i}\alpha_{i+1})]^2\rangle\right.\nonumber\\
&\quad\left.+\left(\LL^\alpha_{33}\right)^2\langle(\Delta \alpha_3)^2\rangle\right\}\nonumber\\
&\geq\mathrm{min}\left(|\LL^x_{11}\LL^p_{11}|+|\LL^x_{21}\LL^p_{21}|+|\LL^x_{22}\LL^p_{22}+\LL^x_{32}\LL^p_{32}|\right.\nonumber\\
&\quad\left.+|\LL^x_{33}\LL^p_{33}|,|\LL^x_{11}\LL^p_{11}+\LL^x_{21}\LL^p_{21}|+|\LL^x_{22}\LL^p_{22}|\right.\nonumber\\
&\quad\left.+|\LL^x_{32}\LL^p_{32}|+|\LL^x_{33}\LL^p_{33}|\right).
\end{align}

At the outset we confirm on a particular example that a fully inseparable but not GME state really does not violate the inequalities (\ref{pcriterion3}) and (\ref{scriterion3}).
As an example we use the family of fully inseparable biseparable states (\ref{eq:rhotestBisep}) with CM $\gamma^{\rm test}$,
Eq.~(\ref{eq:testBisep}), where $r\in(0,1.24)$. For all states with the squeezing parameter $r_{i}=0.05\times i$, $i=0,1,\ldots,24$, we have minimized
numerically the differences $D_{P}$ and $D_{S}$, Eqs.~(\ref{D}) and (\ref{DS}) with $N=3$, i.e., the difference of the LHS and the RHS of inequalities
(\ref{pcriterion3}) and (\ref{scriterion3}). The numerical minimization has been done using a built-in function in Wolfram Mathematica \cite{Mathematica} called \textit{NMinimize} where we specified the used optimization method. On a standard desktop computer, the numerical minimization takes from a few seconds in the case of $N=3$ up to $20$~seconds in the case of $N=6$. Those times are approximately the same as for criteria in \cite{Teh_14} used on the same devices. Our analysis reveals that for all considered values of the squeezing the differences are very small but always
strictly positive numbers, which confirms correctness of the proposed criteria.

Next, we demonstrate the ability of the proposed criteria to detect GME. First, we show this on two sets of basic GME Gaussian states,
including the CV GHZ-like state. Then, we also present two methods allowing us to find numerical examples of Gaussian
states whose GME can be verified by our criteria.

\subsubsection{Analytical examples}

\begin{figure}[htb]
    \centering
    \includegraphics[scale=0.9]{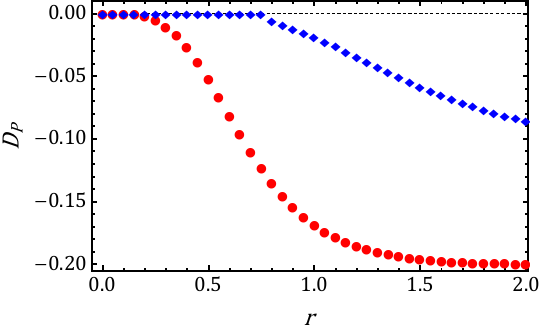}
    \caption{Difference $D_{P}$, Eq.~(\ref{D}), minimized over the elements $\LL^{\alpha}_{ij}\in[-1,1]$ for the CV GHZ-like state with CM (\ref{GHZ}) (red circles) and the
    split squeezed state with CM (\ref{sss}) (blue diamonds) versus the squeezing parameter $r$. See text for details.}
    \label{fig:GHZ}
\end{figure}

Initially, we apply our criteria to the basic Gaussian states, which are widely used for testing of GME criteria.
This includes the family of three-mode CV GHZ-like states with CM (\ref{GHZ}) and the split squeezed states \cite{Teh_22} created by
splitting of a single squeezed state on two balanced beam splitters, which possesses the following CM:
\begin{align}\label{sss}
\gamma^{\rm sss}=\frac{1}{4}&\left[\left(
\begin{array}{ccc}
a_+ & b_+ & -b_+ \\
b_+ & c_+ & d_+ \\
-b_+ & d_+ & c_+
\end{array}\right)\right.\nonumber\\
&\left.\oplus\left(\begin{array}{ccc}
a_- & b_- & -b_- \\
b_- & c_- & d_- \\
-b_- & d_- & c_-
\end{array}
\right)\right],
\end{align}
where
\begin{align}\label{abcd}
a_\pm&=2(1+e^{\pm2r}),\quad &&b_\pm=\sqrt{2}(1-e^{\pm2r}),\nonumber\\
c_\pm&=3+e^{\pm2r},\quad &&d_\pm=1-e^{\pm2r}.
\end{align}
Again, we minimized the difference $D_{P}$, Eq.~(\ref{D}), over the parameters $\LL^{\alpha}_{ij}$. To speed up the numerical search we further imposed on the parameters the following constraint $-1\leq\LL^{\alpha}_{ij}\leq1$. The corresponding minimal differences for the CMs (\ref{GHZ}) and (\ref{sss}) with squeezing parameters $r_{i}=0.05\times i$, $i=0,1,\ldots,40$ are depicted in Fig.~\ref{fig:GHZ}. The figure as well as the numerics reveal that for the GHZ-like state except for the case $r=0$ for all other considered values of the squeezing the minimized difference is negative. Similarly, in the case of the
split squeezed state the minimized difference becomes negative as soon as the considered value of squeezing exceeds the value of $r=0.6$. These two examples clearly demonstrate the ability of
the proposed product criterion (\ref{pcriterion3}) to detect GME of basic three-mode Gaussian states.

Interestingly, the sum criterion (\ref{scriterion3}) fails to detect GME of any of the previous three-mode states
detected by the product criterion. Nevertheless, one can find examples of three-mode GME Gaussian states,
which are detected by both criteria or only by the sum criterion itself.

\subsubsection{Numerical examples}\label{subsec_numerics}

We find examples of Gaussian states whose GME can be detected by the sum criterion (\ref{scriterion3}) using the
Gaussian analog \cite{Nordgren_22} of the iterative algorithm \cite{Paraschiv_17}. Initially, a pure-state
CM $\gamma^{(0)}$ is randomly generated. In the first step of the algorithm an optimal witness $Z^{(1)}$ detecting
GME from minimal set of two-mode marginal CMs of CM $\gamma^{(0)}$ is found. The second step consists of finding
an optimal CM $\gamma^{(1)}$ minimizing $\mbox{Tr}(\gamma Z^{(1)})$. In the next iteration the output CM $\gamma^{(1)}$
is taken as an input to the first step which produces an optimal witness $Z^{(2)}$. The new witness is then used
in the second step to get optimal CM $\gamma^{(2)}$ for the witness, etc. By performing few iterations of the algorithm
on a randomly generated seed CMs we produced one hundred CMs of GME states, which were then tested by our
product and sum GME criteria (\ref{pcriterion3}) and (\ref{scriterion3}). In total, our criteria were capable of detecting GME for nearly 80\% of the CMs.
One of the obtained CMs is given by
\begin{equation}\label{gamma1}
\gamma_{1}=\left(
\begin{array}{ccc}
 6.3 & -6.6 & 2.8 \\
 -6.6 & 10 & -5.6 \\
 2.8 & -5.6 & 5.4 \\
\end{array}
\right)\oplus\left(
\begin{array}{ccc}
 5.4 & 5.3 & 4 \\
 5.3 & 5.5 & 4.2 \\
 4 & 4.2 & 3.5 \\
\end{array}
\right) ,
\end{equation}
where the elements were rounded for brevity to one decimal place in which case physicality of
the rounded CM remained preserved. Minimization of the differences $D_{P}$ and $D_{S}$ over the parameters $\LL^{\alpha}_{ij}\in[-1,1]$
yields $D_{P}=-0.245$ and $D_{S}=-0.165$, where the respective rounded optimal values of the parameters $\LL^{\alpha}_{ij}$
can be found in Appendix~\ref{appendix_sec_II}. As both the differences are negative the GME of the state with CM (\ref{gamma1})
is detected both by the product criterion (\ref{pcriterion3}), as well as sum criterion (\ref{scriterion3}).

Analogously as in the case of CM (\ref{gamma1}) we can find a CM of a Gaussian state whose GME is detected only by the sum criterion.
The CM in question has integer entries and it is of the following form:
\begin{equation}\label{gamma2}
\gamma_{2}=\left(
\begin{array}{ccc}
 5 & -5 & -2 \\
 -5 & 7 & 3 \\
 -2 & 3 & 2 \\
\end{array}
\right)\oplus\left(
\begin{array}{ccc}
 3 & 4 & -3 \\
 4 & 6 & -5 \\
 -3 & -5 & 6 \\
\end{array}
\right).
\end{equation}
Indeed, while the product criterion (\ref{pcriterion3}) yields a non-negative minimal difference $D_{P}$ and thus it does not detect GME,
for the sum criterion (\ref{scriterion3}) one gets $D_{S}=-0.070$, which clearly confirms the presence of GME in the considered state.
The rounded optimal values of the parameters $\LL^{\alpha}_{ij}$ attaining the latter difference are given explicitly in Appendix~\ref{appendix_sec_II}.

\subsubsection{Example derived from a guessed witness}\label{subsec_method}

There is yet another method of how one can find examples of three-mode Gaussian states violating inequality (\ref{scriterion3}). What is more, below we use the method
to find examples of GME states detected by the sum criteria for more than three modes.

Typically, the inseparability criteria based on uncertainty relations are tailored to the state whose entanglement is to be detect, which is known.
Here, on the other hand, we know the criteria and seek states, which are detected by them. This task can be solved relatively easily if we set the parameters
$\LL^{\alpha}_{ij}$ to suitable values. Namely, by using Eq.~(\ref{eq:trace}) and dividing the LHS of the criterion with the RHS, we can express the inequality
(\ref{sumcriterion}) in the form $\mbox{Tr}[\gamma Z]\geq1$, where $Z=Z^{x}\oplus Z^{p}$ and $\gamma=\gamma^{x}\oplus\gamma^{p}$.
Here, $Z$ is a real $2N\times 2N$ symmetric positive-semidefinite matrix and provided that the parameters $\LL^{\alpha}_{ij}$ are chosen properly,
it can detect GME. In fact, the matrix $Z$ is nothing but the already mentioned GME witness in the space of covariance matrices \cite{Hyllus_06}.

In order to learn whether the guessed matrix $Z$ will detect GME, we have to perform its symplectic diagonalization.
Since typically $Z>0$, there is a symplectic matrix $S$, i.e., a real $2N\times 2N$ matrix satisfying the symplectic condition $S\Omega_{N}S^{T}=\Omega_{N}$,
which transforms the matrix as \cite{Williamson_36}
\begin{equation}\label{ZW}
SZS^{T}=\mbox{diag}(\nu_{1},\nu_{2},\ldots,\nu_{N},\nu_{1},\nu_{2},\ldots\nu_{N})\equiv Z_{W}.
\end{equation}
Here, $Z_{W}$ is a diagonal matrix, where the entries $\nu_{i}>0$, $i=1,2,\ldots,N$, are the so-called symplectic eigenvalues.
Now, if $\mbox{Tr}[Z_{W}]<1$ there will be always a CM $\gamma$, for which $\mbox{Tr}(\gamma Z)<1$, which evidences the presence of GME.
To see this, note that $\mbox{Tr}[Z_{W}]=\mbox{Tr}[SZS^{T}]=\mbox{Tr}[S^{T}SZ]$, so if we set $\gamma=S^{T}S$, we have
$\mbox{Tr}[\gamma Z]<1$ and thus the witness matrix $Z$ detects GME carried by the CM $\gamma$.

The GME witness $Z$ can be obtained, e.g., by setting some parameters $\LL^{\alpha}_{ij}$ equal to $+1$ or $-1$, often small parameters,
like $\LL^{\alpha}_{33}$, equal to $10^{-2}$, and by guessing the remaining free parameters. For example, consider the values of the parameters $\LL^{\alpha}_{ij}$ in Tab.~\ref{tabguessedl3}.
\begin{table}[!htb]
\centering
    \caption{Guessed values of parameters $\LL^{\alpha}_{ij}$, $\alpha=x,p$, used to get a GME witness $Z_{3}$.}
    \begin{tabular}{|c|c|c|c|c|}
    \cline{1-2} \cline{4-5}
        $\LL^x_{11}$ & 8 & $\quad$ & $\LL^p_{11}$ & 10 \\ \cline{1-2} \cline{4-5}
        $\LL^x_{21}$ & -1 & $\quad$ & $\LL^p_{21}$ & 2 \\ \cline{1-2} \cline{4-5}
        $\LL^x_{22}$ & 1 & $\quad$ & $\LL^p_{22}$ & 5 \\ \cline{1-2} \cline{4-5}
        $\LL^x_{32}$ & -1 & $\quad$ & $\LL^p_{32}$ & 1 \\ \cline{1-2} \cline{4-5}
        $\LL^x_{33}$ & 0.01 & $\quad$ & $\LL^p_{33}$ & 0.01 \\ \cline{1-2} \cline{4-5}
    \end{tabular}
    \label{tabguessedl3}
\end{table}
Then using Fig.~\ref{fig_RHSconstruction} we get $\mathcal{K}^{(1)}\doteq 86$ and $\mathcal{K}^{(3)}\doteq 84$.
Hence, the witness matrix $Z_{3}=[L^{x}(L^{x})^{\rm T}\oplus L^{p}(L^{p})^{\rm T}]/(2\mathcal{K}^{(3)})$ is
\begin{align}\label{Z3}
    Z_{3}=&\frac{1}{168}\left[\left(\begin{array}{ccc}
        64 & -8 & 0 \\
        -8 & 2 & -1 \\
        0 & -1 & 1.0001
    \end{array}\right)\right.\nonumber\\
    &\quad\left.\oplus\left(\begin{array}{ccc}
        100 & 20 & 0 \\
        20 & 29 & 5 \\
        0 & 5 & 1.0001
    \end{array}\right)\right].
\end{align}
Next, we find numerically the matrix $S_{3}$ symplectically diagonalizing $Z_{3}$, calculate the CM $\gamma_{3}=S_{3}^{T}S_{3}$,
round it for brevity to two decimal places and add to the rounded matrix $10^{-2}\times\openone$ to get a physical CM
(see Appendix~\ref{appendix_subsec_III_N3} for details of the derivation). Thus we arrive at the CM
\begin{align}\label{gamma3}
    \gamma_{3} =& \left(
\begin{array}{ccc}
 1.40 & 1.15 & 0.30 \\
 1.15 & 7.64 & 2.48 \\
 0.30 & 2.48 & 1.49 \\
\end{array}
\right)\nonumber\\
&\oplus\left(
\begin{array}{ccc}
0.84 & -0.15 & 0.09 \\
-0.15 & 0.33 & -0.50 \\
0.09 & -0.50 & 1.49 \\
\end{array}
\right).
\end{align}
To be able to compare the strength of violation of the criterion (\ref{sumcriterion}) by CM $\gamma_{3}$ with previous examples, we again
optimize the difference (\ref{DS}) over $\LL^{\alpha}_{ij}\in[-1,1]$, which yields $D_{S}=-0.15$. This once again confirms that the
proposed sum criterion can really detect GME from minimal set of two-mode marginals.

In the next section we use the latter method to derive Gaussian GME states of $N=4,5$ and $6$ modes, which are detected by the
sum criteria (\ref{sumcriterion}) corresponding to different trees.

\subsubsection{Effect of losses}

In the last part of this section, we discuss the effect of imperfections that may occur in the possible experimental preparation of some of the investigated three-mode states on the detection of the given state by the product criterion (\ref{pcriterion3}). The imperfection we focus on takes the form of the transition of the states through a lossy channel with intensity transmissivity $\eta$.
Our analysis starts with the three-mode CV GHZ-like state with CM (\ref{GHZ}). The best loss robustness for the product criterion (\ref{pcriterion3}) is found for squeezing parameter $r=0.65$ $(-5.65\;\mathrm{dB})$, where minimal transmissivity is approximately $\eta=0.96$ or $4\%$ loss.
Better resilience to losses can be found in the case of the numerically found state with CM (\ref{gamma1}) where the minimal transmissivity is around $\eta=0.92$ or $8\%$ of losses for the form with elements rounded to one decimal place. The original form of CM (\ref{gamma1}) is even more robust with minimal transsitivity $\eta=0.89$ or $11\%$ of losses. Quadrature squeezing required to prepare the state with the CM (\ref{gamma1}) is around $-10\;\mathrm{dB}$. From the experimental perspective it is desirable to look for states requiring smaller amounts of squeezing for the preparation. In that case, we can find another numerical example of a state with CM given in the Appendix~\ref{appendix_sec_IV} that requires only $-4.46\;\mathrm{dB}$ of squeezing. This state can be detected using the presented criteria (\ref{pcriterion3}) starting from a minimal transmissivity of the lossy channel $\eta=0.90$ or $10\%$ of losses.

\subsection{Multi-mode states}

It remains to clarify whether the proposed criteria are also capable to certify GME
in states with more than three modes. Below we use the method of previous section to construct examples of Gaussian states
of up to six modes, whose GME is detected by our criteria. As the method is based on the sum criterion (\ref{sumcriterion}),
the obtained examples violate the sum criterion pronouncedly stronger than the product criterion which often
fails to detect the GME of the constructed state. For this reason, the following discussion is restricted to the sum
criterion (\ref{sumcriterion}).

Initially, we apply the criterion to the four-mode CV GHZ-like state \cite{vanLoock_03}. However, for $r\in(0,2]$,
and both configurations $4a$ and $4b$ (second and third row of Tab.~\ref{table0}), we did not find any set of parameters $\LL^{\alpha}_{ij}$,
for which the sum criterion (\ref{sumcriterion}) (as well as the product criterion (\ref{criterion})) would be violated. This indicates that the
proposed criteria are not suitable for detection of the GME carried by multi-mode GHZ-like states. This can be attributed to the specific
form of GME carried by the GHZ-like states, which manifests itself, as far as the momentum quadratures is concerned, by squeezing in the
global quadrature combination $\sum_{i=1}^{4}p_{i}$, whereas the proposed criteria contain only selected two-mode momentum quadrature
combinations which are for the studied GHZ-like state contaminated by noise.

Nevertheless, using the method of previous section we found examples of GME states of $N=4,5$ and $6$ modes detectable by the sum criterion for
almost all configurations. The exception being criteria corresponding to the five-mode and six-mode linear tree and the six-mode star tree, for
which we so far did not succeed in guessing the GME witness $Z$. The best results obtained by us are summarized in Tab.~\ref{table0}. The table
contains all non-isomorphic trees with up to six vertices, which are rooted and labelled such that the respective sum
criterion contains variances of at most two-mode quadrature combinations. The values of the corresponding difference $D_{S}$, Eq.~(\ref{DS}),
are for given $N$ ordered decreasingly.
\begin{table}[!ht]
\caption{All non-isomorphic trees with $N=3,4,5$ and $6$ vertices. The root of the tree is depicted by a red circle. The linear trees are labeled
according to the standard labeling, whereas the other trees are covered by the reverse level order labeling, which lead to the GME criteria involving variances of at most
two-mode quadrature combinations. The last column contains the difference $D_{S}$, Eq.~(\ref{DS}), for the displayed trees and
CMs constructed by means of the method of Subsec.~\ref{subsec_method}. The values of $D_{S}$ are calculated for the values of parameters
$\LL^{\alpha}_{ij}$ found by a numerical minimization over $\LL^{\alpha}_{ij}\in[-10,10]$. The cases for which we did not found an
example are denoted by a cross. Explicit form of all the CMs as well as other details of derivation of $D_{S}$ can be found in Appendix~\ref{appendix_sec_III}.
See text for details.}
\scalebox{0.92}{\begin{tabular}{|l|lcc|}
        \hline
        $N$ & Index & Labelled tree & Difference $D_{S}$\\ \hline
         &  &  &  \\[-1.7ex]
    3 & $3$ &
        \begin{adjustbox}{valign=c}
            \begin{tikzpicture}[node distance = 0.25 cm]
                \centering
                \tikzset{VertexStyle/.style = {shape          = circle,
                                        thick,
                                        draw            = black,
                                         text           = black,
                                         inner sep      = 0.25pt,
                                         outer sep      = 0pt,
                                         minimum size   = 11 pt}}
                \tikzset{EdgeStyle/.style   = {thin,
                                         double          = black,
                                         double distance = 0.125pt}}
                \node[VertexStyle](A){\footnotesize \textbf{1}};
                \node[VertexStyle,right=of A](B){\footnotesize \textbf{2}};
                \node[circle, thick, draw = red, fill = white, text = black, inner sep= 0.25pt, outer sep = 0pt, minimum size = 11pt,right=of B](C){\footnotesize \color{red} \textbf{3}};
                \draw[EdgeStyle](B) to node{ } (A);
                \draw[EdgeStyle](C) to node{ } (B);
            \end{tikzpicture}
        \end{adjustbox}
        & $-17.87$ \\[1.5 mm]\hline
         &  &  &  \\[-1.7ex]
    4&$4a$ &%
        \begin{adjustbox}{valign=c}
        \begin{tikzpicture}[node distance   = 0.25 cm]
            \centering
          \tikzset{VertexStyle/.style = {shape          = circle,
                                        thick,
                                        draw            = black,
                                         text           = black,
                                         inner sep      = 0.25pt,
                                         outer sep      = 0pt,
                                         minimum size   = 11 pt}}
          \tikzset{EdgeStyle/.style   = {thin,
                                         double          = black,
                                         double distance = 0.125pt}}
             \node[VertexStyle](A){\footnotesize \textbf{1} };
             \node[VertexStyle,right=of A](B){\footnotesize \textbf{2} };
             \node[VertexStyle,right=of B](C){\footnotesize \textbf{3} };
             \node[circle, thick, draw = red, fill = white, text = black, inner sep= 0.25pt, outer sep = 0pt, minimum size = 11pt,right=of C](D){\footnotesize \color{red} \textbf{4} };
             \draw[EdgeStyle](A) to node{ } (B);
             \draw[EdgeStyle](B) to node{ } (C);
             \draw[EdgeStyle](C) to node{ } (D);

          \end{tikzpicture}%
        \end{adjustbox}
                    &   $-20.31$      \\[1.5mm]
   & $4b $&
        \begin{adjustbox}{valign=c}
        \begin{tikzpicture}[node distance   = 0.25 cm]
            \centering
          \tikzset{VertexStyle/.style = {shape          = circle,
                                        thick,
                                        draw            = black,
                                         text           = black,
                                         inner sep      = 0.25pt,
                                         outer sep      = 0pt,
                                         minimum size   = 11 pt}}
          \tikzset{EdgeStyle/.style   = {thin,
                                         double          = black,
                                         double distance = 0.125pt}}
             \node[VertexStyle](A){\footnotesize \textbf{3}};
             \node[circle, thick, draw = red, fill = white, text = black, inner sep= 0.25pt, outer sep = 0pt, minimum size = 11pt,right=of A](B){\footnotesize \color{red} \textbf{4}};
             \node[VertexStyle,right=of B](C){\footnotesize \textbf{1}};
             \node[VertexStyle,below= 0.25 cm of B](D){\footnotesize \textbf{2}};
             \draw[EdgeStyle](B) to node{ } (D);
             \draw[EdgeStyle](A) to node{ } (B);
             \draw[EdgeStyle](B) to node{ } (A);
             \draw[EdgeStyle](B) to node{ } (C);

          \end{tikzpicture}
        \end{adjustbox}
        &   $-5.36$        \\[4.5mm] \hline

         &  &  &  \\[-1.7ex]
         5 & $5a$ &
         \begin{adjustbox}{valign=c}
             \begin{tikzpicture}[node distance = 0.25 cm]
                 \centering
                 \tikzset{VertexStyle/.style = {shape          = circle,
                                        thick,
                                        draw            = black,
                                         text           = black,
                                         inner sep      = 0.25pt,
                                         outer sep      = 0pt,
                                         minimum size   = 11 pt}}
                \tikzset{EdgeStyle/.style   = {thin,
                                         double          = black,
                                         double distance = 0.125pt}}
                    \node[VertexStyle](A){\footnotesize \textbf{1}};
                    \node[VertexStyle, right=of A](B){\footnotesize \textbf{2}};
                    \node[VertexStyle, right=of B](C){\footnotesize \textbf{3}};
                    \node[VertexStyle, right=of C](D){\footnotesize \textbf{4}};
                    \node[circle, thick, draw = red, fill = white, text = black, inner sep= 0.25pt, outer sep = 0pt, minimum size = 11pt, right=of D](E){\footnotesize \color{red} \textbf{5}};
                    \draw[EdgeStyle](A) to node{ } (B);
                    \draw[EdgeStyle](B) to node{ } (C);
                    \draw[EdgeStyle](C) to node{ } (D);
                    \draw[EdgeStyle](D) to node{ } (E);
             \end{tikzpicture}
         \end{adjustbox}
         & \Large{$\times$} \\[2.5 mm]

         & $5b$ &
         \begin{adjustbox}{valign=c}
             \begin{tikzpicture}[node distance = 0.25 cm]
                 \centering
                 \tikzset{VertexStyle/.style = {shape          = circle,
                                        thick,
                                        draw            = black,
                                         text           = black,
                                         inner sep      = 0.25pt,
                                         outer sep      = 0pt,
                                         minimum size   = 11 pt}}
                \tikzset{EdgeStyle/.style   = {thin,
                                         double          = black,
                                         double distance = 0.125pt}}
                    \node[circle, thick, draw = red, fill = white, text = black, inner sep= 0.25pt, outer sep = 0pt, minimum size = 11pt](A){\footnotesize \color{red} \textbf{5}};
                    \node[VertexStyle, right=of A](B){\footnotesize \textbf{2}};
                    \node[VertexStyle, left=of A](C){\footnotesize \textbf{4}};
                    \node[VertexStyle, below=0.25cm of A](D){\footnotesize \textbf{3}};
                    \node[VertexStyle, above=0.25cm of A](E){\footnotesize \textbf{1}};
                    \draw[EdgeStyle](A) to node{ } (B);
                    \draw[EdgeStyle](A) to node{ } (C);
                    \draw[EdgeStyle](A) to node{ } (D);
                    \draw[EdgeStyle](A) to node{ } (E);
             \end{tikzpicture}
         \end{adjustbox}
         & $-3.64$ \\[8 mm]

          & $5c$ &
         \begin{adjustbox}{valign=c}
             \begin{tikzpicture}[node distance = 0.25 cm]
                 \centering
                 \tikzset{VertexStyle/.style = {shape          = circle,
                                        thick,
                                        draw            = black,
                                         text           = black,
                                         inner sep      = 0.25pt,
                                         outer sep      = 0pt,
                                         minimum size   = 11 pt}}
                \tikzset{EdgeStyle/.style   = {thin,
                                         double          = black,
                                         double distance = 0.125pt}}
                    \node[VertexStyle](A){\footnotesize \textbf{1}};
                    \node[VertexStyle, right=of A](B){\footnotesize \textbf{2}};
                    \node[circle, thick, draw = red, fill = white, text = black, inner sep= 0.25pt, outer sep = 0pt, minimum size = 11pt, right=of B](C){\footnotesize \color{red} \textbf{5}};
                    \node[VertexStyle, right=of C](D){\footnotesize \textbf{4}};
                    \node[VertexStyle, below= 0.25cm of C](E){\footnotesize \textbf{3}};
                    \draw[EdgeStyle](A) to node{ } (B);
                    \draw[EdgeStyle](B) to node{ } (C);
                    \draw[EdgeStyle](C) to node{ } (D);
                    \draw[EdgeStyle](C) to node{ } (E);
             \end{tikzpicture}
         \end{adjustbox}
         & $-2.30$ \\[5 mm] \hline
         & & & \\[-1.7ex]

        6  &$6a$&
        \begin{adjustbox}{valign=c}
        \begin{tikzpicture}[node distance   = 0.25 cm]
            \centering
          \tikzset{VertexStyle/.style = {shape          = circle,
                                        thick,
                                        draw            = black,
                                        fill            = white,
                                         text           = black,
                                         inner sep      = 0.25pt,
                                         outer sep      = 0pt,
                                         minimum size   = 11 pt}}
          \tikzset{EdgeStyle/.style   = {thin,
                                         double          = black,
                                         double distance = 0.125pt}}
             \node[VertexStyle](A){\footnotesize \textbf{1} };
             \node[VertexStyle,right=of A](B){\footnotesize \textbf{2} };
             \node[VertexStyle,right=of B](C){\footnotesize \textbf{3} };
             \node[VertexStyle,right=of C](D){\footnotesize \textbf{4} };
             \node[VertexStyle,right=of D](E){\footnotesize \textbf{5} };
             \node[circle, thick, draw = red, fill = white, text = black, inner sep= 0.25pt, outer sep = 0pt, minimum size = 11pt,right=of E](F){\footnotesize \color{red} \textbf{6} };
             \draw[EdgeStyle](A) to node{ } (B);
             \draw[EdgeStyle](B) to node{ } (C);
             \draw[EdgeStyle](C) to node{ } (D);
             \draw[EdgeStyle](D) to node{ } (E);
             \draw[EdgeStyle](E) to node{ } (F);

          \end{tikzpicture}
        \end{adjustbox}
                    &   \Large{$\times$}        \\[2.5mm]

        &$6b$&
        \begin{adjustbox}{valign=c}
        \begin{tikzpicture}[node distance   = 0.25 cm]
            \centering
          \tikzset{VertexStyle/.style = {shape          = circle,
                                        thick,
                                        draw            = black,
                                         text           = black,
                                         inner sep      = 0.25pt,
                                         outer sep      = 0pt,
                                         minimum size   = 11 pt}}
          \tikzset{EdgeStyle/.style   = {thin,
                                         double          = black,
                                         double distance = 0.125pt}}
             \node[circle, thick, draw = red, fill = white, text = black, inner sep= 0.25pt, outer sep = 0pt, minimum size = 11pt](A){\footnotesize \color{red} \textbf{6}};
             \node[VertexStyle,anchor={0}] at (0:0.7cm) {\footnotesize \textbf{3}};
             \node[VertexStyle,anchor={60}] at (60:0.7cm) {\footnotesize \textbf{2}};
             \node[VertexStyle,anchor={120}] at (120:0.7cm) {\footnotesize \textbf{1}};
             \node[VertexStyle,anchor={180}] at (180:0.7cm) {\footnotesize \textbf{5}};
             \node[VertexStyle,anchor={270}] at (270:0.7cm) {\footnotesize \textbf{4}};
             \draw[EdgeStyle](A) to node{ } (0:0.32cm);
             \draw[EdgeStyle](A) to node{ } (60:0.32cm);
             \draw[EdgeStyle](A) to node{ } (120:0.32cm);
             \draw[EdgeStyle](A) to node{ } (180:0.32cm);
             \draw[EdgeStyle](A) to node{ } (270:0.32cm);

          \end{tikzpicture}
        \end{adjustbox}
        &   \Large{$\times$}      \\[6mm]

        &$6c$&
		\begin{adjustbox}{valign=c}
        \begin{tikzpicture}[node distance   = 0.25 cm]
            \centering
          \tikzset{VertexStyle/.style = {shape          = circle,
                                        thick,
                                        draw            = black,
                                         text           = black,
                                         inner sep      = 0.25pt,
                                         outer sep      = 0pt,
                                         minimum size   = 11 pt}}
          \tikzset{EdgeStyle/.style   = {thin,
                                         double          = black,
                                         double distance = 0.125pt}}
             \node[VertexStyle](A){\footnotesize \textbf{2} };
             \node[VertexStyle,right=of A](B){\footnotesize \textbf{5} };
             \node[circle, thick, draw = red, fill = white, text = black, inner sep= 0.25pt, outer sep = 0pt, minimum size = 11pt,right=of B](C){\footnotesize \color{red} \textbf{6} };
             \node[VertexStyle,right=of C](D){\footnotesize \textbf{3} };
             \node[VertexStyle,right=of D](E){\footnotesize \textbf{1} };
             \node[VertexStyle,below= 0.25 cm of C](F){\footnotesize \textbf{4} };
             \draw[EdgeStyle](C) to node{ } (F);
             \draw[EdgeStyle](A) to node{ } (B);
             \draw[EdgeStyle](B) to node{ } (C);
             \draw[EdgeStyle](C) to node{ } (D);
             \draw[EdgeStyle](D) to node{ } (E);

          \end{tikzpicture}
        \end{adjustbox}
                    &   $-1.00$         \\[4.5mm]

        &$6d$&
        \begin{adjustbox}{valign=c}
        \begin{tikzpicture}[node distance   = 0.25 cm]
            \centering
          \tikzset{VertexStyle/.style = {shape          = circle,
                                        thick,
                                        draw            = black,
                                         text           = black,
                                         inner sep      = 0.25pt,
                                         outer sep      = 0pt,
                                         minimum size   = 11 pt}}
          \tikzset{EdgeStyle/.style   = {thin,
                                         double          = black,
                                         double distance = 0.125pt}}
             \node[VertexStyle](A){\footnotesize \textbf{3} };
             \node[VertexStyle,right=of A](B){\footnotesize \textbf{5} };
             \node[circle, thick, draw = red, fill = white, text = black, inner sep= 0.25pt, outer sep = 0pt, minimum size = 11pt,right=of B](C){\footnotesize \color{red} \textbf{6} };
             \node[VertexStyle,right=of C](D){\footnotesize \textbf{4} };
             \node[VertexStyle,right=of D](E){\footnotesize \textbf{1} };
             \node[VertexStyle,below= 0.25 cm of D](F){\footnotesize \textbf{2} };
             \draw[EdgeStyle](D) to node{ } (F);
             \draw[EdgeStyle](A) to node{ } (B);
             \draw[EdgeStyle](B) to node{ } (C);
             \draw[EdgeStyle](C) to node{ } (D);
             \draw[EdgeStyle](D) to node{ } (E);

          \end{tikzpicture}
        \end{adjustbox}
                    &   $-0.91$         \\[4.5mm]

        &$6e$&
        \begin{adjustbox}{valign=c}
        \begin{tikzpicture}[node distance   = 0.25 cm]
            \centering
          \tikzset{VertexStyle/.style = {shape          = circle,
                                        thick,
                                        draw            = black,
                                         text           = black,
                                         inner sep      = 0.25pt,
                                         outer sep      = 0pt,
                                         minimum size   = 11 pt}}
          \tikzset{EdgeStyle/.style   = {thin,
                                         double          = black,
                                         double distance = 0.125pt}}
             \node[](A){ };
             \node[VertexStyle,right=of A](B){\footnotesize \textbf{5} };
             \node[circle, thick, draw = red, fill = white, text = black, inner sep= 0.25pt, outer sep = 0pt, minimum size = 11pt,right=of B](C){\footnotesize \color{red} \textbf{6} };
             \node[,right=of C](D){ };
             \node[VertexStyle,below= 0.1 cm of A](E){\footnotesize \textbf{1} };
             \node[VertexStyle,above= 0.1 cm of A](F){\footnotesize \textbf{2} };
             \node[VertexStyle,below= 0.1 cm of D](G){\footnotesize \textbf{4} };
             \node[VertexStyle,above= 0.1 cm of D](H){\footnotesize \textbf{3} };
             \draw[EdgeStyle](B) to node{ } (E);
             \draw[EdgeStyle](B) to node{ } (F);
             \draw[EdgeStyle](C) to node{ } (G);
             \draw[EdgeStyle](C) to node{ } (H);
             \draw[EdgeStyle](B) to node{ } (C);

          \end{tikzpicture}
        \end{adjustbox}
                    &   $-0.65$         \\[5mm]

        &$6f$&
        \begin{adjustbox}{valign=c}
        \begin{tikzpicture}[node distance   = 0.25 cm]
            \centering
          \tikzset{VertexStyle/.style = {shape          = circle,
                                        thick,
                                        draw            = black,
                                         text           = black,
                                         inner sep      = 0.25pt,
                                         outer sep      = 0pt,
                                         minimum size   = 11 pt}}
          \tikzset{EdgeStyle/.style   = {thin,
                                         double          = black,
                                         double distance = 0.125pt}}
             \node[VertexStyle](A){\footnotesize \textbf{5} };
              \node[circle, thick, draw = red, fill = white, text = black, inner sep= 0.25pt, outer sep = 0pt, minimum size = 11pt,right=of A](C){\footnotesize  \color{red} \textbf{6} };
             \node[VertexStyle,right=of C](B){\footnotesize \textbf{4} };
             \node[VertexStyle,right=of B](D){\footnotesize \textbf{2} };
             \node[VertexStyle,below= 0.25 cm of B](E){\footnotesize \textbf{3} };
             \node[VertexStyle,above= 0.25 cm of B](F){\footnotesize \textbf{1} };
             \draw[EdgeStyle](B) to node{ } (E);
             \draw[EdgeStyle](B) to node{ } (F);
             \draw[EdgeStyle](A) to node{ } (C);
             \draw[EdgeStyle](B) to node{ } (C);
             \draw[EdgeStyle](B) to node{ } (D);

          \end{tikzpicture}
        \end{adjustbox}
                    &   $-0.13$         \\[8mm] \hline
\end{tabular}}
\label{table0}
\end{table}
Since optimization over $\LL^{\alpha}_{ij}\in[-1,1]$ yields typically for $N>3$ small values of $D_{S}$, the values of the difference in Tab.~\ref{table0} were obtained
for optimization over an extended interval $\LL^{\alpha}_{ij}\in[-10,10]$. This is why the value of $D_{S}$ for the three-mode case is roughly
hundred-times larger compared to the examples presented in previous section. The Tab.~\ref{table0} indicates that up to the configurations $3$ and $4a$
the value of $D_{S}$ decreases with increasing $N$, which is a typical behaviour of GME states (cf. Tab.~I of Ref.~\cite{Paraschiv_17}).
The exception is likely to be attributed to the fact that in the case $3$ we guessed a little worse witness than for configuration $4a$.
All the guessed witnesses, the CMs of the states, and other details of derivation of the results in Tab.~\ref{table0} are given explicitly
in Appendix~\ref{appendix_sec_III}.

We accomplished the present analysis by an independent check of the presence of GME verifiable only from the minimal set of
two-mode marginal CMs in the Gaussian states used in Tab.~ \ref{table0}. For this purpose, we used the already mentioned method of entanglement
witnesses in the space of CMs \cite{Hyllus_06}, which detect the GME from the minimal set of two-mode marginal CMs \cite{Nordgren_22}.
More precisely, for each CM and all considered configurations we found numerically such a witness, which independently confirms
the presence of the GME detectable only from the respective minimal set of marginals in the constructed examples.

\section{Conclusions}\label{sec_conclusions}

In this paper we derived a set of product and sum criteria for GME of continuous-variable systems. The criteria are based on the second moments of
quadrature operators and unlike the overwhelming majority of the known GME criteria contain only the variances of at most two-mode
quadrature combinations. Moreover, the presented criteria are also minimal in the sense that they contain only the minimum number
of the latter variances. In other words, the criteria require only the knowledge of the minimal set of two-mode marginal CMs, which consists of $7N-4$ independent CM elements of the state under study \cite{Nordgren_22}. Since our criteria do not take into account the $x-p$ correlations, the number of the elements further reduces to $4N-2$ and thus it scales only linearly with the number of modes. This distinguishes our criteria from the most commonly used GME criteria \cite{vanLoock_03, Teh_14, Toscano_15}, which require knowledge of the entire CM without $x-p$ correlations, i.e., knowledge of its $N(N+1)$ elements. Other known $N$-mode CV GME criteria \cite{Hyllus_06} utilize the whole CM including $x-p$ correlations, i.e., to use them one needs to know its $N(2N+1)$ elements. All these criteria thus scale quadratically with the number of modes.
Since any minimal set of marginal CMs can be represented by a special kind of a graph known as a tree, each criterion is intimately associated to a particular tree. This allowed us to develop a method of construction of all the sum criteria directly
from the underlying tree.

Further, we have shown that the product criterion detects GME of the three-mode CV GHZ-like state and the split squeezed state,
but neither of the states is detected by the sum criterion. The fact that we detected GME in the three-mode CV GHZ-like state from its two-mode marginals is in contrast with results for qubit GHZ state where this is not possible \cite{Gittsovich_10}. We have also constructed numerical examples of three-mode Gaussian GME states which are
simultaneously detected by both the criteria as well as only by the sum criterion. To get examples of GME states detected by our criteria for
more than three modes, we developed a method based on symplectic diagonalization of a guessed witness in the space of CMs. Thus we
found states detected by the sum criterion for nearly all configurations of up to six modes.

Interestingly, our criteria do not detect the GME in three- and four-mode GME states with all two-mode marginals separable, which were recently found in
Ref.~\cite{Nordgren_22}. Therefore, we tested the criteria for another more than hundred numerically generated GME states with separable marginals
and we always got a negative result. This rises a question of how to modify the present criteria so as to make them capable of detecting the GME
from separable marginals, which is deferred for further research. Likewise, inspired by the existence of the CV GME criteria for higher-order
moments \cite{Shchukin_14,Zhang_23} we could try to generalize the proposed criteria to the higher-order moments, which is thus another
open problem. Finally, one may ask whether the present algorithm can be used to
derive analogous minimal criteria for genuine multipartite quantum steering \cite{He_13,Teh_22}. Since the steering criteria are typically obtained only by
small modifications of derivation of the standard GME criteria \cite{Teh_14,Armstrong_15}, we expect that the same approach will be
viable also in the case of our criteria, which will be addressed elsewhere.

The present paper provides a rigorous method of construction of minimal criteria of GME. The method relies on the concept of witness matrix in the
space of CMs and graph theory, and gives a recipe of how to systematically construct similar minimal criteria for other types of multipartite
quantum correlations. The GME criteria presented here are also experimentally friendly as they are simple and immediately reveal how to measure them.
What is more, the criteria do not require to measure entire CM but only its strictly smaller part which is needed for detection
of GME. This feature provides an advantage compared to other GME criteria by saving the number of measurements needed for testing of GME of the investigated state.
We believe that our results will stimulate theoretical research of minimal criteria of multipartite quantum properties and find applications
in characterization of the properties in experimentally prepared quantum states.

\acknowledgments
O.L. acknowledges support from IGA-PrF-2023-006 and the European Union’s 2020 research and
innovation program (CSA - Coordination and support action, H2020-WIDESPREAD-2020-5) under grant agreement No. 951737 (NONGAUSS). O.L. and L.M. acknowledge the project 8C22002 (CVStar) of MEYS of the Czech Republic, which has received funding from the European Union's Horizon 2020 research and innovation programme under the Grant Agreements no. 731473 and 101017733; and the project  'Quantum Secure Networks Partnership' (QSNP), which has received funding from the European Union's Horizon Europe research and innovation programme under the Grant Agreement no. 101114043.

\section*{Author contribution}
\noindent
Both authors contributed equally to the results and the manuscript writing.

\appendix

\section{Fully inseparable biseparable Gaussian state}\label{appendix_sec_I}

\begin{figure}[htb]
    \centering
    \includegraphics[scale=0.88]{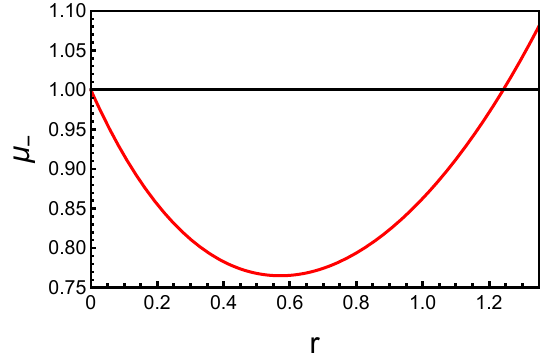}
    \caption{Lower symplectic eigenvalue $\mu_{-}$, Eq.~(\ref{mu}), versus the squeezing parameter $r$ for the state with CM $\gamma^{\rm test}_{13}$, Eq.~(\ref{gamma13}).}
    \label{fig:mu}
\end{figure}

This section is dedicated to the proof that the three-mode biseparable state (\ref{eq:rhotestBisep}) with CM (\ref{eq:testBisep}) is in a certain region of squeezing parameters fully inseparable, i.e., it is entangled
across all three bipartite splits $1|23, 2|31$ and $3|12$. The state possesses a fully symmetric CM of the form
\begin{equation}\label{gammatest}
    \gamma^{\rm test}=\begin{pmatrix}
b & e & e\\
e & b & e\\
e & e & b
\end{pmatrix}\oplus\begin{pmatrix}
b & -e & -e\\
-e & b & -e\\
-e & -e & b
\end{pmatrix},
\end{equation}
where
\begin{equation}
b=\frac{2\cosh(2r)+1}{3},\quad e=\frac{\sinh(2r)}{3}.
\end{equation}
Due to the symmetry it is sufficient to verify entanglement of the state with respect to just one split, e.g., $3|12$, which we will do using the results of Ref.~\cite{Fiurasek_07,Vidal_02}.
Interference of modes $1$ and $2$ on a balanced beam splitter does not change entanglement properties across the split and decouples mode $2$ from modes $1$ and $3$. Hence, we are
left with the CM
\begin{equation}\label{gamma13}
    \gamma^{\rm test}_{13}=\begin{pmatrix}
b+e & \sqrt{2}e \\
\sqrt{2}e & b
\end{pmatrix}\oplus\begin{pmatrix}
b-e & -\sqrt{2}e \\
-\sqrt{2}e & b
\end{pmatrix}
\end{equation}
of modes $1$ and $3$. Entanglement of the state with the latter CM then can be proved easily by using positive partial transposition criterion \cite{Simon_00} expressed in terms of the symplectic eigenvalues \cite{Vidal_02}. The partial transposition operation with respect to mode $3$ transforms the CM (\ref{gamma13}) as $\gamma_{13}^{{\rm test}\,(T_{3})}=(\openone\oplus\sigma_{z})\gamma_{13}^{\rm test}(\openone\oplus\sigma_{z})$ and its symplectic eigenvalues $\mu_{\pm}$ are obtained from the spectrum $\{\pm i\mu_{+},\pm i\mu_{-}\}$ of the matrix $\Omega_{2}\gamma_{13}^{{\rm test}\,(T_{3})}$ \cite{Vidal_02} in the form
\begin{equation}\label{mu}
\mu_{\pm}=\sqrt{\frac{\alpha\pm\sqrt{\alpha^{2}-4\beta}}{2}},
\end{equation}
where
$\alpha=2b^2+3e^2$ and $\beta=(b^2-2e^2)^2-(be)^2$. According to the partial transposition criterion the state with CM (\ref{gamma13}) is entangled
if the minimal symplectic eigenvalue $\mu_{-}$ satisfies $\mu_{-}<1$. Thus we are looking for an interval of
squeezing parameters $r\geq0$, for which $\mu_{-}<1$. In Fig.~\ref{fig:mu} we plot the dependence of the symplectic eigenvalue
$\mu_{-}$ on the squeezing parameter $r$.
The figure shows that as $r$ increases from $r=0$ when $\mu_{-}=1$ the
symplectic eigenvalue satisfies $\mu_{-}<1$ for up to $r=r_{\rm max}$ when $\mu_{-}=1$ and then $\mu_{-}>1$ with further increase of
$r$. The threshold value of $r_{\rm max}$ can be calculated from the condition $\mu_{-}=1$,
which simplifies to $\alpha-\beta=1$ and leads further to the following equation:
\begin{equation}
3e^{12r}-34e^{10r}-35e^{8r}+132e^{6r}-35e^{4r}-34e^{2 r}+3=0.
\end{equation}
Upon solving the latter equation numerically one finally finds $r_{\rm max}\doteq1.24$ and thus the state with CM (\ref{gamma13}) is entangled for $r\in(0,1.24)$.
As a consequence, the state with CM (\ref{gammatest}) is fully inseparable for $r\in(0,1.24)$.

\section{Derivation of the structure of the matrix $Z_{A}^{\alpha}$ providing two-mode quadrature combinations $u_{i}^{\alpha}$}\label{appendix_sec_new}

In this section we show what structure the matrix $Z_{A}^{\alpha}$ must have for the operators $u_{i}^{x,p}$, $i=1,2,\ldots,N-1$, Eq.~(\ref{uxpi}) of the main text, to contain at most two non-zero terms. In the main text, we showed that for operators $u_{i}^{x,p}$ to contain at most two non-zero terms, the vectors $L^{\alpha}_{i+1:N,i}$, $i=1,2,\ldots,N-1$, Eq.~(\ref{Livect}) of the main text, have to possess at most one non-zero component. Below we show that the vectors $L^{\alpha}_{i+1:N,i}$, will have at most one non-zero component if the vectors $(Z_{A}^{\alpha})_{i+1:N,i}$, $i=1,2,\ldots,N-1$, contain at most one non-zero component.

Namely, for a positive-definite matrix $Z_{A}^{\alpha}$ the vectors $L^{\alpha}_{i+1:N,i}$ can be expressed as \cite[Theorem~4.2.5]{Golub_96}
\begin{equation}\label{L1}
L^{\alpha}_{2:N,1}=\frac{(Z_{A}^{\alpha})_{2:N,1}}{L_{11}^{\alpha}}
\end{equation}
and
\begin{equation}\label{Li}
L^{\alpha}_{i+1:N,i}=\frac{1}{L_{ii}^{\alpha}}\left[(Z_{A}^{\alpha})_{i+1:N,i}-\sum_{j=1}^{i-1}L_{ij}^{\alpha}L^{\alpha}_{i+1:N,j}\right]
\end{equation}
for $i=2,3,\ldots,N-1$. From Eq.~(\ref{L1}) it then follows that the vector $L^{\alpha}_{2:N,1}$ will contain at most one non-zero component if
the vector $(Z_{A}^{\alpha})_{2:N,1}$ has at most one non-zero component. Moving to the vector $L^{\alpha}_{3:N,2}$, Eq.~(\ref{Li}) with $i=2$, we see that if
the vector $(Z_{A}^{\alpha})_{2:N,1}$ has at most one non-zero component, then the second term in the square brackets
vanishes because it contains products of different components of the vector $(Z_{A}^{\alpha})_{2:N,1}$, and $L^{\alpha}_{3:N,2}=(Z_{A}^{\alpha})_{3:N,2}/L^{\alpha}_{22}$
according to Eq.~(\ref{Li}). Again, in order the vector $L^{\alpha}_{3:N,2}$ to have at most one non-zero component,
the vector $(Z_{A}^{\alpha})_{3:N,2}$ has to possess at most one non-zero component. Repeating the same procedure $(i-1)$-times we see that the vectors
$L^{\alpha}_{j+1:N,j}$, $j=1,2,\ldots,i-1$ will have at most one non-zero component if the same holds for vectors
$(Z_{A}^{\alpha})_{j+1:N,j}$, $j=1,2,\ldots,i-1$. If this is the case, then the sum $\sum_{j=1}^{i-1}L_{ij}^{\alpha}L^{\alpha}_{i+1:N,j}$
in Eq.~(\ref{Li}) vanishes, because each term $L_{ij}^{\alpha}L^{\alpha}_{i+1:N,j}$ is a vector with components given by
a product of {\it different} components of the vector $(Z_{A}^{\alpha})_{j+1:N,j}$. The $i$-th vector (\ref{Li}) of the Cholesky matrix
$L^{\alpha}$ then reduces to
\begin{equation}\label{Li2appendix}
L^{\alpha}_{i+1:N,i}=\frac{(Z_{A}^{\alpha})_{i+1:N,i}}{L_{ii}^{\alpha}}
\end{equation}
and again it has at most one non-zero component if the vector $(Z_{A}^{\alpha})_{i+1:N,i}$ contains at most one non-zero component.

Summarizing the obtained results we see that if we require all the quadrature combinations $u_{i}^{x,p}$, Eq.~(\ref{uxpi}) of the main text, to contain at most two non-zero terms, then all vectors $(Z_{A}^{\alpha})_{i+1:N,i}$, $i=1,2,\ldots,N-1$ have to possess at most one non-zero component, as we wanted to prove.

\section{Optimal parameters}\label{appendix_sec_II}

This section contains optimal values of the parameters $\LL^{\alpha}_{ij}$ minimizing the differences (\ref{D}) and (\ref{DS}) between the
LHS and the RHS of the proposed minimal GME criteria.

For the three-mode CM (\ref{gamma1}) and the product criterion (\ref{pcriterion3}) the optimal parameters $\LL^{\alpha}_{ij}$ are summarized in Tab.~\ref{tabgamma1P}
\begin{table}[!htb]
\centering
    \caption{Rounded values of the parameters $\LL^{\alpha}_{ij}\in[-1,1]$, $\alpha=x,p$, used to calculate the difference $D_{P}=-0.245$ for the product criterion (\ref{pcriterion3}) and the CM
    $\gamma_{1}$, Eq.~(\ref{gamma1}).}
    \begin{tabular}{|c|c|c|c|c|}
    \cline{1-2} \cline{4-5}
        $\LL^x_{11}$ & 0.947 & $\quad$ & $\LL^p_{11}$ & 0.989 \\ \cline{1-2} \cline{4-5}
        $\LL^x_{21}$ & 0.760 & $\quad$ & $\LL^p_{21}$ & -0.987 \\ \cline{1-2} \cline{4-5}
        $\LL^x_{22}$ & 0.860 & $\quad$ & $\LL^p_{22}$ & 0.817 \\ \cline{1-2} \cline{4-5}
        $\LL^x_{32}$ & 1 & $\quad$ & $\LL^p_{32}$ & -1 \\ \cline{1-2} \cline{4-5}
        $\LL^x_{33}$ & $1\times10^{-5}$ & $\quad$ & $\LL^p_{33}$ & $1.8\times10^{-6}$ \\ \cline{1-2} \cline{4-5}
    \end{tabular}
    \label{tabgamma1P}
\end{table}

In the case of CM (\ref{gamma1}) and the sum criterion (\ref{scriterion3}) the optimal parameters $\LL^{\alpha}_{ij}$ are summarized in Tab.~\ref{tabgamma1S}
\begin{table}[!htb]
\centering
    \caption{Rounded values of the parameters $\LL^{\alpha}_{ij}\in[-1,1]$, $\alpha=x,p$, used to calculate the difference $D_{S}=-0.165$ for the sum criterion (\ref{scriterion3}) and the CM
    $\gamma_{1}$, Eq.~(\ref{gamma1}).}
    \begin{tabular}{|c|c|c|c|c|}
    \cline{1-2} \cline{4-5}
        $\LL^x_{11}$ & 0.425  & $\quad$ & $\LL^p_{11}$ & 1 \\ \cline{1-2} \cline{4-5}
        $\LL^x_{21}$ & 0.253 & $\quad$ & $\LL^p_{21}$ & -0.954 \\ \cline{1-2} \cline{4-5}
        $\LL^x_{22}$ & 0.305 & $\quad$ & $\LL^p_{22}$ & 0.792 \\ \cline{1-2} \cline{4-5}
        $\LL^x_{32}$ & 0.505 & $\quad$ & $\LL^p_{32}$ & -1 \\ \cline{1-2} \cline{4-5}
        $\LL^x_{33}$ & $2\times10^{-4}$ & $\quad$ & $\LL^p_{33}$ & $4\times10^{-4}$ \\ \cline{1-2} \cline{4-5}
    \end{tabular}
    \label{tabgamma1S}
\end{table}

Optimal parameters $\LL^{\alpha}_{ij}$ for the CM (\ref{gamma2}) and the sum criterion (\ref{scriterion3}) are summarized in Tab.~\ref{tabgamma2S}
\begin{table}[!htb]
\centering
    \caption{Rounded values of the parameters $\LL^{\alpha}_{ij}\in[-1,1]$, $\alpha=x,p$, used to calculate the difference $D_{S}=-0.070$ for the sum criterion (\ref{scriterion3}) and the CM
    $\gamma_{2}$, Eq.~(\ref{gamma2}).}
    \begin{tabular}{|c|c|c|c|c|}
    \cline{1-2} \cline{4-5}
        $\LL^x_{11}$ & 0.514  & $\quad$ & $\LL^p_{11}$ & 1 \\ \cline{1-2} \cline{4-5}
        $\LL^x_{21}$ & 0.314  & $\quad$ & $\LL^p_{21}$ & -0.652 \\ \cline{1-2} \cline{4-5}
        $\LL^x_{22}$ & 0.419  & $\quad$ & $\LL^p_{22}$ & 0.489 \\ \cline{1-2} \cline{4-5}
        $\LL^x_{32}$ & -0.908  & $\quad$ & $\LL^p_{32}$ & 0.559 \\ \cline{1-2} \cline{4-5}
        $\LL^x_{33}$ & $3\times10^{-5}$ & $\quad$ & $\LL^p_{33}$ & $8.7\times10^{-7}$ \\ \cline{1-2} \cline{4-5}
    \end{tabular}
    \label{tabgamma2S}
\end{table}

\section{Derivation of examples from a guessed witness}\label{appendix_sec_III}

In this section we present a complete derivation of Gaussian states of $N\geq3$ modes violating minimal GME
criteria (\ref{criterion}) and (\ref{sumcriterion}), based on a guessed witness in the space of CMs.

\subsection{N=3}\label{appendix_subsec_III_N3}

We start with the guessed three-mode witness matrix $Z_{3}$, Eq.~(\ref{Z3}).
Making use of the method \cite{Pereira_21}, we can find numerically the matrix $S_{3}$ symplectically diagonalizing $Z_{3}$ in the form

\begin{equation}\label{S3}
S_{3}=\begin{pmatrix}
\mathbb{O}_{3} & X_{3} \\
Y_{3} & \mathbb{O}_{3}
\end{pmatrix},
\end{equation}
where
\begin{equation}\label{X3}
    X_{3}=\left(\begin{array}{ccc}
        -0.0445813 & 0.222904 & -1.11455\\
        0.0879785 & -0.503613 &  0.492602\\
        0.904559 & -0.110546 & -0.00252406
         \end{array}\right)
\end{equation}
and
\begin{equation}\label{Y3}
    Y_{3}=\left(\begin{array}{ccc}
      -0.139315   & -1.11452 & -1.11455\\
      -0.309427    & -2.52072 & -0.491755 \\
       1.12874  & 0.190238 & -0.00710211
         \end{array}\right).
\end{equation}
From the symplectic matrix (\ref{S3}) we then get the following CM:

\begin{widetext}
\begin{equation}\label{gamma3p}
    \gamma_{3}'=S_{3}^{T}S_{3}=\left(\begin{array}{ccc}
       1.38922 & 1.14998  &  0.299419\\
      1.14998   & 7.63236  &  2.48042 \\
      0.299419   & 2.48042  & 1.4841
    \end{array}\right)\oplus\left(\begin{array}{ccc}
     0.827955    & -0.15424 &  0.0907433\\
       -0.15424   & 0.315532  & -0.49624 \\
      0.0907433   & -0.49624  & 1.48488
    \end{array}\right).
\end{equation}
\end{widetext}
Next, for simplicity we round all elements of the latter CM to two decimal places thereby getting the matrix $\bar{\gamma}_{3}$,
which slightly violates the Heisenberg uncertainty principle $\gamma+i\Omega_{3}\geq0$ \cite{Simon_00}, Eq.~(\ref{ineq:Heisenbergg})
with $N=3$ of the main text. Therefore, we have added to the CM a small amount of thermal noise thereby creating a physical
CM $\gamma_{3}=\bar{\gamma}_{3}+0.01\times\openone$ given explicitly in Eq.~(\ref{gamma3}).
Now, if we optimize numerically the difference (\ref{DS}) for the sum criterion (\ref{scriterion3}) over the parameters $\LL^{\alpha}_{ij}\in[-1,1]$,
we arrive at $D_{S}=-0.15$ and the optimal parameters are listed in Tab.~\ref{tabgamma3S}.
\begin{table}[!htb]
\centering
    \caption{Rounded values of the parameters $\LL^{\alpha}_{ij}\in[-1,1]$, $\alpha=x,p$, used to calculate the difference (\ref{DS}) for the sum criterion (\ref{scriterion3}) and the CM
    $\gamma_{3}$, Eq.~(\ref{gamma3}).}
    \begin{tabular}{|c|c|c|c|c|}
    \cline{1-2} \cline{4-5}
        $\LL^x_{11}$ & 0.90  & $\quad$ & $\LL^p_{11}$ & 1 \\ \cline{1-2} \cline{4-5}
        $\LL^x_{21}$ & -0.23  & $\quad$ & $\LL^p_{21}$ & 0.98 \\ \cline{1-2} \cline{4-5}
        $\LL^x_{22}$ & 0.23  & $\quad$ & $\LL^p_{22}$ & 1 \\ \cline{1-2} \cline{4-5}
        $\LL^x_{32}$ &  -1  & $\quad$ & $\LL^p_{32}$ & 1 \\ \cline{1-2} \cline{4-5}
        $\LL^x_{33}$ & 0.01  & $\quad$ & $\LL^p_{33}$ & 0.01 \\ \cline{1-2} \cline{4-5}
    \end{tabular}
    \label{tabgamma3S}
\end{table}

The difference (\ref{DS}) in the first row of Tab.~\ref{table0} of the main text has been calculated for the unrounded CM $\gamma_{3}'$, Eq.~(\ref{gamma3p}).
If we optimize numerically the difference (\ref{DS}) for the sum criterion (\ref{scriterion3}) over the parameters $\LL^{\alpha}_{ij}\in[-10,10]$,
we arrive at $D_{S}=-17.87$ and the optimal parameters are listed in Tab.~\ref{tabgamma3S10}.
\begin{table}[!htb]
\centering
    \caption{Rounded values of the parameters $\LL^{\alpha}_{ij}\in[-10,10]$, $\alpha=x,p$, used to calculate the difference $D_{S}$, Eq.~(\ref{DS}), given in the first row
    of Tab.~\ref{table0} for the CM $\gamma_{3}'$, Eq.~(\ref{gamma3p}).}
    \begin{tabular}{|c|c|c|c|c|}
    \cline{1-2} \cline{4-5}
        $\LL^x_{11}$ & 9.11  & $\quad$ & $\LL^p_{11}$ & 10 \\ \cline{1-2} \cline{4-5}
        $\LL^x_{21}$ & -2.31  & $\quad$ & $\LL^p_{21}$ & 10 \\ \cline{1-2} \cline{4-5}
        $\LL^x_{22}$ & 2.31  & $\quad$ & $\LL^p_{22}$ & 10 \\ \cline{1-2} \cline{4-5}
        $\LL^x_{32}$ &  -10  & $\quad$ & $\LL^p_{32}$ & 10 \\ \cline{1-2} \cline{4-5}
        $\LL^x_{33}$ & 0.01  & $\quad$ & $\LL^p_{33}$ & 0.01 \\ \cline{1-2} \cline{4-5}
    \end{tabular}
    \label{tabgamma3S10}
\end{table}

\subsection{N=4}\label{appendix_subsec_III_N4}

\subsubsection{Linear tree 4a}\label{appendix_subsec_III_N4a}

We start with the guessed four-mode witness matrix $Z_{4a}$ of the linear tree $4a$,
\begin{equation}\label{Z4a}
    Z_{4a}=\frac{1}{5.5002}\left(\begin{array}{cc}
        Z_{4a}^{x} & \mathbb{0}_{4} \\
        \mathbb{0}_{4} & Z_{4a}^{p}
    \end{array}\right),
\end{equation}
where
\begin{equation}
    Z_{4a}^{x}=\left(
\begin{array}{cccc}
 1 & -0.5 & 0 & 0 \\
 -0.5 & 0.5 & -0.5 & 0 \\
 0 & -0.5 & 1.25 & -0.5 \\
 0 & 0 & -0.5 & 1.0001 \\
\end{array}
\right)\nonumber\\
\end{equation}
and
\begin{equation}
    Z_{4a}^{p}=\left(
\begin{array}{cccc}
 1 & 0.5 & 0 & 0 \\
 0.5 & 0.5 & 0.5 & 0 \\
 0 & 0.5 & 1.25 & 0.5 \\
 0 & 0 & 0.5 & 1.0001 \\
\end{array}
\right).\nonumber\\
\end{equation}
Next, we find numerically the matrix $S_{4a}$ symplectically diagonalizing $Z_{4a}$ in the form:

\begin{equation}\label{S4a}
S_{4a}=\begin{pmatrix}
\mathbb{O}_{4} & X_{4a} \\
Y_{4a} & \mathbb{O}_{4}
\end{pmatrix},
\end{equation}
where
\begin{widetext}
\begin{equation}\label{X4a}
    X_{4a}=\left(
\begin{array}{cccc}
 -0.666646 & 1.33331 & -0.666635 & 0.333288 \\
 -1.05853 & 0.895417 & -0.870291 & 0.275906 \\
 0.0468749 & -0.178567 & -0.191418 & 1.00347 \\
 0.571074 & 0.0886146 & -0.850668 & 0.204715 \\
\end{array}
\right)
\end{equation}
and
\begin{equation}\label{Y4a}
    Y_{4a}=\left(
\begin{array}{cccc}
 -0.666646 & -1.33331 & -0.666635 & -0.333288 \\
 1.05853 & 0.895417 & 0.870291 & 0.275906 \\
 0.0468749 & 0.178567 & -0.191418 & -1.00347 \\
 -0.571074 & 0.0886146 & 0.850668 & 0.204715 \\
\end{array}
\right).
\end{equation}
From the symplectic matrix (\ref{S4a}) we then get the following CM:


\begin{align}\label{gamma4a}
    \gamma_{4a}=S_{4a}^{T}S_{4a}=&\left(
\begin{array}{cccc}
 1.89322 & 1.79444 & 0.87087 & 0.350295 \\
 1.79444 & 2.61924 & 1.70931 & 0.530383 \\
 0.87087 & 1.70931 & 1.96209 & 0.828527 \\
 0.350295 & 0.530383 & 0.828527 & 1.23606 \\
\end{array}
\right)\nonumber\\
&\oplus\left(
\begin{array}{cccc}
 1.89322 & -1.79444 & 0.87087 & -0.350295 \\
 -1.79444 & 2.61924 & -1.70931 & 0.530383 \\
 0.87087 & -1.70931 & 1.96209 & -0.828527 \\
 -0.350295 & 0.530383 & -0.828527 & 1.23606 \\
\end{array}
\right).
\end{align}
\end{widetext}
Now, if we optimize numerically the difference (\ref{DS}) for the sum criterion (\ref{sumcriterion}) over the parameters $\LL^{\alpha}_{ij}\in[-10,10]$,
we arrive at $D_{S}=-20.31$ and the optimal parameters are listed in Tab.~\ref{tabgamma4aS}.
\begin{table}[!htb]
\centering
    \caption{Rounded values of the parameters $\LL^{\alpha}_{ij}\in[-10,10]$, $\alpha=x,p$, used to calculate the difference (\ref{DS}) for the sum criterion (\ref{sumcriterion}) and the CM
    $\gamma_{4a}$, Eq.~(\ref{gamma4a}).}
    \begin{tabular}{|c|c|c|c|c|}
    \cline{1-2} \cline{4-5}
        $\LL^x_{11}$ & 10  & $\quad$ & $\LL^p_{11}$ & 10 \\ \cline{1-2} \cline{4-5}
        $\LL^x_{21}$ & -5.08  & $\quad$ & $\LL^p_{21}$ & 5.08 \\ \cline{1-2} \cline{4-5}
        $\LL^x_{22}$ & 5.09  & $\quad$ & $\LL^p_{22}$ & 5.08 \\ \cline{1-2} \cline{4-5}
        $\LL^x_{32}$ & -9.03  & $\quad$ & $\LL^p_{32}$ & -9.03 \\ \cline{1-2} \cline{4-5}
        $\LL^x_{33}$ & 5.07  & $\quad$ & $\LL^p_{33}$ & 5.10 \\ \cline{1-2} \cline{4-5}
        $\LL^x_{43}$ & -10 & $\quad$ & $\LL^p_{43}$ & 10 \\ \cline{1-2} \cline{4-5}
        $\LL^x_{44}$ & 0.01 & $\quad$ & $\LL^p_{44}$ & 0.01 \\ \cline{1-2} \cline{4-5}
    \end{tabular}
    \label{tabgamma4aS}
\end{table}

\subsubsection{Star tree 4b}\label{appendix_subsec_III_N4b}

We guess the four-mode witness matrix $Z_{4b}$ corresponding to the star tree $4b$ in the following form:
\begin{equation}\label{Z4b}
    Z_{4b}=\frac{1}{68.02}\left(\begin{array}{cc}
        Z_{4b}^{x} & \mathbb{0}_{4} \\
        \mathbb{0}_{4} & Z_{4b}^{p}
    \end{array}\right),
\end{equation}
where
\begin{equation*}
    Z_{4b}^{x}=\left(
\begin{array}{cccc}
 4 & 0 & 0 & -2 \\
 0 & 1 & 0 & -1 \\
 0 & 0 & 4 & -2 \\
 -2 & -1 & -2 & 3.01 \\
\end{array}
\right)\nonumber\\
\end{equation*}
and
\begin{equation*}
    Z_{4b}^{p}=\left(
\begin{array}{cccc}
 100 & 0 & 0 & 10 \\
 0 & 25 & 0 & 5 \\
 0 & 0 & 25 & 5 \\
 10 & 5 & 5 & 3.01 \\
\end{array}
\right).\nonumber\\
\end{equation*}
The matrix $S_{4b}$ symplectically diagonalizing $Z_{4b}$ is of the form:

\begin{equation}\label{S4b}
S_{4b}=\begin{pmatrix}
V_{4b} & X_{4b} \\
Y_{4b} & W_{4b}
\end{pmatrix},
\end{equation}
where
\vspace{0.1mm}
\begin{widetext}
\begin{equation}\label{V4b}
    V_{4b}=\left(
\begin{array}{cccc}
 0 & 0 & 0 & 0 \\
 0 & 0 & 0 & 0 \\
 0.258041 & -0.339431 & 1.61423 & 0.280456 \\
 0 & 0 & 0 & 0 \\
\end{array}
\right),
\end{equation}
\begin{equation}\label{X4b}
    X_{4b}=\left(
\begin{array}{cccc}
 -0.123786 & -0.247003 & -0.247382 & 1.23881 \\
 -0.0795257 & -0.506179 & -0.206184 & 0.647292 \\
 0 & 0 & 0 & 0 \\
 0.454241 & -0.0135947 & -0.0420425 & -0.192404 \\
\end{array}
\right),
\end{equation}
\begin{equation}\label{Y4b}
    Y_{4b}=\left(
\begin{array}{cccc}
 0.61893 & 1.23502 & 0.618456 & 1.23881 \\
 -0.397629 & -2.53089 & -0.515461 & -0.647292 \\
 0 & 0 & 0 & 0 \\
 2.27121 & -0.0679731 & -0.105106 & 0.192405 \\
\end{array}
\right),
\end{equation}
and
\begin{equation}\label{W4b}
    W_{4b}=\left(
\begin{array}{cccc}
 0 & 0 & 0 & 0 \\
 0 & 0 & 0 & 0 \\
 0.051608 & -0.067886 & 0.645693 & -0.280457 \\
 0 & 0 & 0 & 0 \\
\end{array}
\right).
\end{equation}
From the symplectic matrix (\ref{S4b}) we then get the following CM:

\begin{align}\label{gamma4b}
    \gamma_{4b}=S_{4b}^{T}S_{4b}=&\left(
\begin{array}{cccc}
 5.76616 & 1.52878 & 0.765563 & 1.53348 \\
 1.52878 & 8.05052 & 1.5276 & 3.0599 \\
 0.765563 & 1.5276 & 3.26498 & 1.5323 \\
 1.53348 & 3.0599 & 1.5323 & 2.06932 \\
\end{array}
\right)\nonumber\\
&\oplus\left(
\begin{array}{cccc}
 0.230645 & 0.061151 & 0.0612449 & -0.306695 \\
 0.061151 & 0.322021 & 0.122208 & -0.611981 \\
 0.0612449 & 0.122208 & 0.522397 & -0.612921 \\
 -0.306695 & -0.611981 & -0.612921 & 2.06932 \\
\end{array}
\right).
\end{align}
\end{widetext}
Now, if we optimize numerically the difference (\ref{DS}) for the sum criterion (\ref{sumcriterion}) over the parameters $\LL^{\alpha}_{ij}\in[-10,10]$,
we arrive at $D_{S}=-5.36$ and the optimal parameters are listed in Tab.~\ref{tabgamma4aS}.
\begin{table}[!htb]
\centering
    \caption{Rounded values of the parameters $\LL^{\alpha}_{ij}\in[-10,10]$, $\alpha=x,p$, used to calculate the difference (\ref{DS}) for the sum criterion (\ref{sumcriterion}) and the CM
    $\gamma_{4b}$, Eq.~(\ref{gamma4b}).}
    \begin{tabular}{|c|c|c|c|c|}
    \cline{1-2} \cline{4-5}
        $\LL^x_{11}$ & 2.35  & $\quad$ & $\LL^p_{11}$ & 10 \\ \cline{1-2} \cline{4-5}
        $\LL^x_{41}$ & -2.27  & $\quad$ & $\LL^p_{41}$ & 2.13 \\ \cline{1-2} \cline{4-5}
        $\LL^x_{22}$ & 2.17  & $\quad$ & $\LL^p_{22}$ & 10 \\ \cline{1-2} \cline{4-5}
        $\LL^x_{42}$ & -2.42  & $\quad$ & $\LL^p_{42}$ & 2.00 \\ \cline{1-2} \cline{4-5}
        $\LL^x_{33}$ & 4.09  & $\quad$ & $\LL^p_{33}$ & 9.99 \\ \cline{1-2} \cline{4-5}
        $\LL^x_{43}$ & -2.28 & $\quad$ & $\LL^p_{43}$ & 2.13 \\ \cline{1-2} \cline{4-5}
        $\LL^x_{44}$ & 0.12 & $\quad$ & $\LL^p_{44}$ & 0.09 \\ \cline{1-2} \cline{4-5}
    \end{tabular}
    \label{tabgamma4bS}
\end{table}

\subsection{N=5}\label{appendix_subsec_III_N5}

\subsubsection{Star tree 5b}\label{appendix_subsec_III_N5b}

We guess the five-mode witness matrix $Z_{5b}$ corresponding to the star tree $5b$,
\newpage
\begin{equation}\label{Z5b}
    Z_{5b}=\frac{1}{2.4801}\left(\begin{array}{cc}
        Z_{5b}^{x} & \mathbb{0}_{5} \\
        \mathbb{0}_{5} & Z_{5b}^{p}
    \end{array}\right),
\end{equation}
where
\begin{equation*}
    Z_{5b}^{x}=\left(
\begin{array}{ccccc}
 1 & 0 & 0 & 0 & -0.1 \\
 0 & 0.25 & 0 & 0 & -0.05 \\
 0 & 0 & 1 & 0 & -0.1 \\
 0 & 0 & 0 & 0.25 & -0.05 \\
 -0.1 & -0.05 & -0.1 & -0.05 & 0.0401 \\
\end{array}
\right)
\end{equation*}
and
\begin{equation*}
    Z_{5b}^{p}=\left(
\begin{array}{ccccc}
 1 & 0 & 0 & 0 & 0.1 \\
 0 & 0.25 & 0 & 0 & 0.05 \\
 0 & 0 & 1 & 0 & 0.1 \\
 0 & 0 & 0 & 0.25 & 0.05 \\
 0.1 & 0.05 & 0.1 & 0.05 & 0.0401 \\
\end{array}
\right).
\end{equation*}
By finding numerically the matrix $S_{5b}$ symplectically diagonalizing $Z_{5b}$, we then get the following CM:
\begin{equation}\label{gamma5b}
    \gamma_{5b}=S_{5b}^TS_{5b}=\left(\begin{array}{cc}
        \gamma_{5b}^{x} & \mathbb{0}_{5} \\
        \mathbb{0}_{5} & \gamma_{5b}^{p}
    \end{array}\right),
\end{equation}
where
\vspace{0.3cm}
\begin{widetext}
\begin{equation}\label{gamma5bX}
    \gamma_{5b}^{x}=\left(
\begin{array}{ccccc}
 1.02221 & 0.0444158 & 0.0222153 & 0.0444158 & 0.222178 \\
 0.0444158 & 1.0888 & 0.0444158 & 0.0888027 & 0.444208 \\
 0.0222153 & 0.0444158 & 1.02221 & 0.0444158 & 0.222178 \\
 0.0444158 & 0.0888027 & 0.0444158 & 1.0888 & 0.444208 \\
 0.222178 & 0.444208 & 0.222178 & 0.444208 & 1.22203 \\
\end{array}
\right),
\end{equation}
\begin{equation}\label{gamma5bP}
    \gamma_{5b}^{p}=\left(
\begin{array}{ccccc}
 1.02221 & 0.0444158 & 0.0222153 & 0.0444158 & -0.222178 \\
 0.0444158 & 1.0888 & 0.0444158 & 0.0888027 & -0.444208 \\
 0.0222153 & 0.0444158 & 1.02221 & 0.0444158 & -0.222178 \\
 0.0444158 & 0.0888027 & 0.0444158 & 1.0888 & -0.444208 \\
 -0.222178 & -0.444208 & -0.222178 & -0.444208 & 1.22203 \\
\end{array}
\right).
\end{equation}
\end{widetext}
Now, if we optimize numerically the difference (\ref{DS}) for the sum criterion (\ref{sumcriterion}) over the parameters $\LL^{\alpha}_{ij}\in[-10,10]$,
we arrive at $D_{S}=-3.64$.

\subsubsection{Tree 5c}\label{appendix_subsec_III_N5c}

We start with the guessed five-mode witness matrix $Z_{5c}$ corresponding to the tree $5c$,
\begin{equation}\label{Z5c}
    Z_{5c}=\frac{1}{2.4801}\left(\begin{array}{cc}
        Z_{5c}^{x} & \mathbb{0}_{5} \\
        \mathbb{0}_{5} & Z_{5c}^{p}
    \end{array}\right),
\end{equation}
where
\newpage
\begin{equation}
    Z_{5c}^{x}=\left(
\begin{array}{ccccc}
 1 & -0.1 & 0 & 0 & 0 \\
 -0.1 & 0.26 & 0 & 0 & -0.05 \\
 0 & 0 & 1 & 0 & -0.1 \\
 0 & 0 & 0 & 0.25 & -0.05 \\
 0 & -0.05 & -0.1 & -0.05 & 0.0301 \\
\end{array}
\right)\nonumber\\
\end{equation}
and
\begin{equation}
    Z_{5c}^{p}=\left(
\begin{array}{ccccc}
 1 & 0.1 & 0 & 0 & 0 \\
 0.1 & 0.26 & 0 & 0 & 0.05 \\
 0 & 0 & 1 & 0 & 0.1 \\
 0 & 0 & 0 & 0.25 & 0.05 \\
 0 & 0.05 & 0.1 & 0.05 & 0.0301 \\
\end{array}
\right).\nonumber\\
\end{equation}
Next, we find numerically the matrix $S_{5c}$ symplectically diagonalizing $Z_{5c}$ and from that symplectic matrix we then get the following CM:
\begin{equation}\label{gamma5c}
    \gamma_{5c}=S_{5c}^{T}S_{5c}=\left(\begin{array}{cc}
        \gamma_{5c}^{x} & \mathbb{0}_{5} \\
        \mathbb{0}_{5} & \gamma_{5c}^{p}
    \end{array}\right),
\end{equation}
where

\vspace{2mm}
\begin{widetext}
\begin{equation}\label{gamma5cX}
    \gamma_{5c}^{x}=\left(
\begin{array}{ccccc}
 1.01364 & 0.16921 & 0.00397589 & 0.00844686 & 0.035641 \\
 0.16921 & 1.10058 & 0.0438767 & 0.0877648 & 0.438485 \\
 0.00397589 & 0.0438767 & 1.02196 & 0.0439097 & 0.219645 \\
 0.00844686 & 0.0877648 & 0.0439097 & 1.08779 & 0.439145 \\
 0.035641 & 0.438485 & 0.219645 & 0.439145 & 1.19671 \\
\end{array}
\right),
\end{equation}
\begin{equation}\label{gamma5cP}
    \gamma_{5c}^{p}=\left(
\begin{array}{ccccc}
 1.01364 & -0.16921 & -0.00397589 & -0.00844686 & 0.035641 \\
 -0.16921 & 1.10058 & 0.0438767 & 0.0877648 & -0.438485 \\
 -0.00397589 & 0.0438767 & 1.02196 & 0.0439097 & -0.219645 \\
 -0.00844686 & 0.0877648 & 0.0439097 & 1.08779 & -0.439145 \\
 0.035641 & -0.438485 & -0.219645 & -0.439145 & 1.19671 \\
\end{array}
\right).
\end{equation}
\end{widetext}
Now, if we optimize numerically the difference (\ref{DS}) for the sum criterion (\ref{sumcriterion}) over the parameters $\LL^{\alpha}_{ij}\in[-10,10]$,
we arrive at $D_{S}=-2.30$.


\subsection{N=6}\label{appendix_subsec_III_N6}

\subsubsection{Tree 6c}\label{appendix_subsec_III_N6c}

The guessed six-mode witness matrix $Z_{6c}$ for the tree $6c$ is of the form:
\begin{equation}\label{Z6c}
    Z_{6c}=\frac{1}{3.2301}\left(\begin{array}{cc}
        Z_{6c}^{x} & \mathbb{0}_{6} \\
        \mathbb{0}_{6} & Z_{6c}^{p}
    \end{array}\right),
\end{equation}
where
\begin{widetext}
\begin{equation*}
    Z_{6c}^{x}=\left(
\begin{array}{cccccc}
 1 & 0 & -0.1 & 0 & 0 & 0 \\
 0 & 1 & 0 & 0 & -0.1 & 0 \\
 -0.1 & 0 & 1.01 & 0 & 0 & -0.1 \\
 0 & 0 & 0 & 0.25 & 0 & -0.05 \\
 0 & -0.1 & 0 & 0 & 0.02 & -0.01 \\
 0 & 0 & -0.1 & -0.05 & -0.01 & 0.0301 \\
\end{array}
\right)
\end{equation*}
\end{widetext}
and
\begin{equation*}
    Z_{6c}^{p}=\left(
\begin{array}{cccccc}
 1 & 0 & 0.1 & 0 & 0 & 0 \\
 0 & 1 & 0 & 0 & 0.1 & 0 \\
 0.1 & 0 & 1.01 & 0 & 0 & 0.1 \\
 0 & 0 & 0 & 0.25 & 0 & 0.05 \\
 0 & 0.1 & 0 & 0 & 0.02 & 0.01 \\
 0 & 0 & 0.1 & 0.05 & 0.01 & 0.0301 \\
\end{array}
\right).
\end{equation*}
From the numerically calculated matrix $S_{6c}$ symplectically diagonalizing $Z_{6c}$ we then get the following CM:
\begin{equation}\label{gamma6c}
    \gamma_{6c}=S_{6c}^TS_{6c}=\left(\begin{array}{cc}
        \gamma_{6c}^{x} & \mathbb{0}_{6} \\
        \mathbb{0}_{6} & \gamma_{6c}^{p}
    \end{array}\right),
\end{equation}
where
\begin{widetext}
\begin{equation}\label{gamma6cX}
    \gamma_{6c}^{x}=\left(
\begin{array}{cccccc}
 1.00601 & 0.00996299 & 0.110751 & 0.0220412 & 0.0996522 & 0.102581 \\
 0.00996299 & 1.1082 & 0.0996528 & 0.198639 & 1.0828 & 0.99682 \\
 0.110751 & 0.0996528 & 1.11751 & 0.224225 & 0.996235 & 1.12667 \\
 0.0220412 & 0.198639 & 0.224225 & 1.44671 & 1.98457 & 2.2454 \\
 0.0996522 & 1.0828 & 0.996235 & 1.98457 & 9.83666 & 9.97538 \\
 0.102581 & 0.99682 & 1.12667 & 2.2454 & 9.97538 & 10.2866 \\
\end{array}
\right),
\end{equation}
\begin{equation}\label{gamma6cP}
    \gamma_{6c}^{p}=\left(
\begin{array}{cccccc}
 1.00601 & 0.00996299 & -0.110751 & -0.0220412 & -0.0996522 & 0.102581 \\
 0.00996299 & 1.1082 & -0.0996528 & -0.198639 & -1.0828 & 0.99682 \\
 -0.110751 & -0.0996528 & 1.11751 & 0.224225 & 0.996235 & -1.12667 \\
 -0.0220412 & -0.198639 & 0.224225 & 1.44671 & 1.98457 & -2.2454 \\
 -0.0996522 & -1.0828 & 0.996235 & 1.98457 & 9.83666 & -9.97538 \\
 0.102581 & 0.99682 & -1.12667 & -2.2454 & -9.97538 & 10.2866 \\
\end{array}
\right).
\end{equation}
\end{widetext}
Now, if we optimize numerically the difference (\ref{DS}) for the sum criterion (\ref{sumcriterion}) over the parameters $\LL^{\alpha}_{ij}\in[-10,10]$,
we arrive at $D_{S}=-1.00$.

\subsubsection{Tree 6d}\label{appendix_subsec_III_N6d}

For the tree $6d$ we guessed the witness matrix $Z_{6d}$ as follows:
\begin{equation}\label{Z6d}
    Z_{6d}=\frac{1}{2.7301}\left(\begin{array}{cc}
        Z_{6d}^{x} & \mathbb{0}_{6} \\
        \mathbb{0}_{6} & Z_{6d}^{p}
    \end{array}\right),
\end{equation}
where
\begin{widetext}
\begin{equation*}
    Z_{6d}^{x}=\left(
\begin{array}{cccccc}
 1 & 0 & 0 & -0.1 & 0 & 0 \\
 0 & 1 & 0 & -0.1 & 0 & 0 \\
 0 & 0 & 1 & 0 & -0.1 & 0 \\
 -0.1 & -0.1 & 0 & 0.27 & 0 & -0.05 \\
 0 & 0 & -0.1 & 0 & 0.02 & -0.01 \\
 0 & 0 & 0 & -0.05 & -0.01 & 0.0201 \\
\end{array}
\right)
\end{equation*}
\end{widetext}
and
\begin{equation*}
    Z_{6d}^{p}=\left(
\begin{array}{cccccc}
 1 & 0 & 0 & 0.1 & 0 & 0 \\
 0 & 0.25 & 0 & 0.05 & 0 & 0 \\
 0 & 0 & 1 & 0 & 0.1 & 0 \\
 0.1 & 0.05 & 0 & 0.27 & 0 & 0.05 \\
 0 & 0 & 0.1 & 0 & 0.02 & 0.01 \\
 0 & 0 & 0 & 0.05 & 0.01 & 0.0201 \\
\end{array}
\right).
\end{equation*}
In the next step, we find numerically the matrix $S_{6d}$ symplectically diagonalizing $Z_{6d}$, which yields
\begin{equation}\label{gamma6d}
    \gamma_{6d}=S_{6d}^{T}S_{6d}=\left(\begin{array}{cc}
        \gamma_{6d}^{x} & \mathbb{0}_{6} \\
        \mathbb{0}_{6} & \gamma_{6d}^{p}
    \end{array}\right),
\end{equation}
where
\begin{widetext}
\begin{equation}\label{gamma6dX}
    \gamma_{6d}^{x}=\left(
\begin{array}{cccccc}
 1.01727 & 0.0150336 & 0.0200197 & 0.206136 & 0.200098 & 0.21641 \\
 0.0150336 & 0.513163 & 0.0200209 & 0.178237 & 0.200154 & 0.211026 \\
 0.0200197 & 0.0200209 & 1.10944 & 0.200099 & 1.09521 & 1.00374 \\
 0.206136 & 0.178237 & 0.200099 & 1.47744 & 1.99935 & 2.24369 \\
 0.200098 & 0.200154 & 1.09521 & 1.99935 & 9.96135 & 10.0452 \\
 0.21641 & 0.211026 & 1.00374 & 2.24369 & 10.0452 & 10.2858 \\
\end{array}
\right),
\end{equation}
\begin{equation}\label{gamma6dP}
    \gamma_{6d}^{p}=\left(
\begin{array}{cccccc}
 1.01727 & 0.0300672 & 0.0200197 & -0.206136 & -0.200098 & 0.21641 \\
 0.0300672 & 2.05265 & 0.0400418 & -0.356474 & -0.400308 & 0.422052 \\
 0.0200197 & 0.0400418 & 1.10944 & -0.200099 & -1.09521 & 1.00374 \\
 -0.206136 & -0.356474 & -0.200099 & 1.47744 & 1.99935 & -2.24369 \\
 -0.200098 & -0.400308 & -1.09521 & 1.99935 & 9.96135 & -10.0452 \\
 0.21641 & 0.422052 & 1.00374 & -2.24369 & -10.0452 & 10.2858 \\
\end{array}
\right).
\end{equation}
\end{widetext}
Now, if we optimize numerically the difference (\ref{DS}) for the sum criterion (\ref{sumcriterion}) over the parameters $\LL^{\alpha}_{ij}\in[-10,10]$,
we arrive at $D_{S}=-0.91$.

\subsubsection{Tree 6e}\label{appendix_subsec_III_N6e}

In the case of tree $6e$ we guessed the witness matrix $Z_{6e}$ as
\begin{equation}\label{Z6e}
    Z_{6e}=\frac{1}{3.1801}\left(\begin{array}{cc}
        Z_{6e}^{x} & \mathbb{0}_{6} \\
        \mathbb{0}_{6} & Z_{6e}^{p}
    \end{array}\right),
\end{equation}
where
\begin{widetext}
\begin{equation*}
    Z_{6e}^{x}=\left(
\begin{array}{cccccc}
 1 & 0 & 0 & 0 & -0.1 & 0 \\
 0 & 1 & 0 & 0 & -0.1 & 0 \\
 0 & 0 & 0.01 & 0 & 0 & -0.01 \\
 0 & 0 & 0 & 0.01 & 0 & -0.01 \\
 -0.1 & -0.1 & 0 & 0 & 0.03 & -0.01 \\
 0 & 0 & -0.01 & -0.01 & -0.01 & 0.0301 \\
\end{array}
\right)
\end{equation*}
and
\begin{equation*}
    Z_{6e}^{p}=\left(
\begin{array}{cccccc}
 4 & 0 & 0 & 0 & 0.2 & 0 \\
 0 & 1 & 0 & 0 & 0.1 & 0 \\
 0 & 0 & 1 & 0 & 0 & 0.1 \\
 0 & 0 & 0 & 1 & 0 & 0.1 \\
 0.2 & 0.1 & 0 & 0 & 0.03 & 0.01 \\
 0 & 0 & 0.1 & 0.1 & 0.01 & 0.0301 \\
\end{array}
\right).\nonumber\\
\end{equation*}
\end{widetext}
From the numerically found symplectic matrix $S_{6e}$ symplectically diagonalizing $Z_{6e}$ we then construct the following CM:
\begin{equation}\label{gamma6e}
    \gamma_{6e}=S_{6e}^{T}S_{6e}=\left(\begin{array}{cc}
        \gamma_{6e}^{x} & \mathbb{0}_{6} \\
        \mathbb{0}_{6} & \gamma_{6e}^{p}
    \end{array}\right),
\end{equation}
where

\vspace{0.2cm}

\begin{widetext}
\begin{equation}\label{gamma6eX}
    \gamma_{6e}^{x}=\left(
\begin{array}{cccccc}
 2.07887 & 0.0788541 & 0.729928 & 0.729928 & 0.788929 & 0.749595 \\
 0.0788541 & 1.07884 & 0.730827 & 0.730827 & 0.788736 & 0.749621 \\
 0.729928 & 0.730827 & 19.0655 & 9.06554 & 7.28942 & 9.32218 \\
 0.729928 & 0.730827 & 9.06554 & 19.0655 & 7.28942 & 9.32218 \\
 0.788929 & 0.788736 & 7.28942 & 7.28942 & 6.89127 & 7.49556 \\
 0.749595 & 0.749621 & 9.32218 & 9.32218 & 7.49556 & 8.5861 \\
\end{array}
\right),
\end{equation}
\begin{equation}\label{gamma6eP}
    \gamma_{6e}^{p}=\left(
\begin{array}{cccccc}
 0.519719 & 0.0394271 & -0.0364964 & -0.0364964 & -0.394465 & 0.374798 \\
 0.0394271 & 1.07884 & -0.0730827 & -0.0730827 & -0.788736 & 0.749621 \\
 -0.0364964 & -0.0730827 & 0.190655 & 0.0906554 & 0.728942 & -0.932218 \\
 -0.0364964 & -0.0730827 & 0.0906554 & 0.190655 & 0.728942 & -0.932218 \\
 -0.394465 & -0.788736 & 0.728942 & 0.728942 & 6.89127 & -7.49556 \\
 0.374798 & 0.749621 & -0.932218 & -0.932218 & -7.49556 & 8.5861 \\
\end{array}
\right).
\end{equation}
\end{widetext}
Now, if we optimize numerically the difference (\ref{DS}) for the sum criterion (\ref{sumcriterion}) over the parameters $\LL^{\alpha}_{ij}\in[-10,10]$,
we arrive at $D_{S}=-0.65$.

\subsubsection{Tree 6f}\label{appendix_subsec_III_N6f}

In the case of the tree $6f$ we guess the six-mode witness matrix $Z_{6f}$ to be
\begin{equation}\label{Z6f}
    Z_{6f}=\frac{1}{75.4801}\left(\begin{array}{cc}
        Z_{6f}^{x} & \mathbb{0}_{6} \\
        \mathbb{0}_{6} & Z_{6f}^{p}
    \end{array}\right),
\end{equation}
where
\begin{widetext}
\begin{equation*}
    Z_{6f}^{x}=\left(
\begin{array}{cccccc}
 25 & 0 & 0 & -0.5 & 0 & 0 \\
 0 & 25 & 0 & -0.5 & 0 & 0 \\
 0 & 0 & 25 & -0.5 & 0 & 0 \\
 -0.5 & -0.5 & -0.5 & 0.04 & 0. & -0.01 \\
 0 & 0 & 0 & 0 & 25 & -0.5 \\
 0 & 0 & 0 & -0.01 & -0.5 & 0.0201 \\
\end{array}
\right)
\end{equation*}
\end{widetext}
and
\begin{equation*}
    Z_{6f}^{p}=\left(
\begin{array}{cccccc}
 25 & 0 & 0 & 5 & 0 & 0 \\
 0 & 25 & 0 & 0.5 & 0 & 0 \\
 0 & 0 & 25 & 0.5 & 0 & 0 \\
 5 & 0.5 & 0.5 & 2.02 & 0 & 0.1 \\
 0 & 0 & 0 & 0 & 0.01 & 0.01 \\
 0 & 0 & 0 & 0.1 & 0.01 & 0.0201 \\
\end{array}
\right).
\end{equation*}
From the symplectic matrix $S_{6f}$ symplectically diagonalizing $Z_{6f}$ we get the CM of the following form:
\begin{equation}\label{gamma6f}
    \gamma_{6f}=S_{6f}^{T}S_{6f}=\left(\begin{array}{cc}
        \gamma_{6f}^{x} & \mathbb{0}_{6} \\
        \mathbb{0}_{6} & \gamma_{6f}^{p}
    \end{array}\right),
\end{equation}
where
\begin{widetext}
\begin{equation}\label{gamma6fX}
    \gamma_{6f}^{x}=\left(
\begin{array}{cccccc}
 1.00888 & 0.0070959 & 0.0070959 & 0.44554 & 0.000907054 & 0.045205 \\
 0.0070959 & 1.00531 & 0.00530564 & 0.265838 & 0.00090708 & 0.0452774 \\
 0.0070959 & 0.00530564 & 1.00531 & 0.265838 & 0.00090708 & 0.0452774 \\
 0.44554 & 0.265838 & 0.265838 & 12.3541 & 0.0453557 & 2.26827 \\
 0.000907054 & 0.00090708 & 0.00090708 & 0.0453557 & 0.020908 & 0.0454074 \\
 0.045205 & 0.0452774 & 0.0452774 & 2.26827 & 0.0454074 & 1.27089 \\
\end{array}
\right),
\end{equation}
\begin{equation}\label{gamma6fP}
    \gamma_{6f}^{p}=\left(
\begin{array}{cccccc}
 1.00889 & 0.00267199 & 0.00267199 & -0.0446333 & -0.0453539 & 0.0452056 \\
 0.00267199 & 1.00044 & 0.000437932 & -0.0224336 & -0.00453415 & 0.00444848 \\
 0.00267199 & 0.000437932 & 1.00044 & -0.0224336 & -0.00453415 & 0.00444848 \\
 -0.0446333 & -0.0224336 & -0.0224336 & 0.124334 & 0.226785 & -0.226828 \\
 -0.0453539 & -0.00453415 & -0.00453415 & 0.226785 & 52.2702 & -2.27039 \\
 0.0452056 & 0.00444848 & 0.00444848 & -0.226828 & -2.27039 & 1.27089 \\
\end{array}
\right).
\end{equation}
\end{widetext}

Now, if we optimize numerically the difference (\ref{DS}) for the sum criterion (\ref{sumcriterion}) over the parameters $\LL^{\alpha}_{ij}\in[-10,10]$,
we arrive at $D_{S}=-0.13$.

\section{Numerical example of robust GME state}\label{appendix_sec_IV}

This section contains CM of state found by Gaussian analog \cite{Nordgren_22} of the iterative algorithm \cite{Paraschiv_17}, as is done in Sec. \ref{subsec_numerics}. In contrast to the states with CMs $\gamma_1$ and $\gamma_2$ presented there, we used additional constraint restricting the amount of squeezing needed to prepare the state with given CM to maximally $-5\;\mathrm{dB}$. The state with CM
\begin{align}\label{gamma7}
    \gamma_7=&\left(\begin{array}{ccc}
    3.96432 & 3.33486 & 1.66686 \\
    3.33486 & 3.60552 & 1.84641 \\
    1.66686 & 1.84641 & 1.27596
\end{array}\right)\nonumber\\
&\oplus\left(\begin{array}{ccc}
    3.48332 & -2.61754 & 3.45572 \\
    -2.61754 & 3.13354 & -3.379668 \\
    3.45572 & -3.79668 & 7.73756
\end{array}\right)
\end{align}
is detected as GME by criterion (\ref{pcriterion3}) with the difference $D_P=-0.111$ over the parameters $\LL^\alpha_{ij}\in[-1,1]$ given in Tab. \ref{tabgamma7}.
\begin{table}[t]
    \centering
    \caption{Rounded values of the parameters $\LL^{\alpha}_{ij}\in[-1,1]$, $\alpha=x,p$, used to calculate the difference (\ref{D}) for the product criterion (\ref{pcriterion3}) and the CM $\gamma_{7}$, Eq.~(\ref{gamma7}).}
    \begin{tabular}{|c|c|c|c|c|}
    \cline{1-2} \cline{4-5}
        $\LL^x_{11}$ & 0.67 & $\quad$ & $\LL^p_{11}$ & 0.82 \\\cline{1-2} \cline{4-5}
        $\LL^x_{12}$ & -0.76 & $\quad$ & $\LL^p_{12}$ & 1.00 \\\cline{1-2} \cline{4-5}
        $\LL^x_{22}$ & 0.68 & $\quad$ & $\LL^p_{22}$ & 1.00 \\\cline{1-2} \cline{4-5}
        $\LL^x_{23}$ & -1.00 & $\quad$ & $\LL^p_{23}$ & 0.61 \\\cline{1-2} \cline{4-5}
        $\LL^x_{33}$ & 0.01 & $\quad$ & $\LL^p_{33}$ & 0.01 \\\cline{1-2} \cline{4-5}
    \end{tabular}
    \label{tabgamma7}
\end{table}


\bibliographystyle{quantum}
\bibliography{Accepted/biblio}

\end{document}